\documentclass[a4paper,10pt]{article}
\usepackage{latexsym}
\usepackage{amsmath}
\usepackage{amssymb}
\usepackage{amscd}

\newlength{\minitwocolumn}
\newcommand{\bz}{{\mathbb Z}}

\newcommand{\bc}{{\mathbb C}}

\makeatletter
\@addtoreset{equation}{section}
\makeatother

\newtheorem{thm}{Theorem}[section]
\newtheorem{prop}[thm]{Proposition}
\newtheorem{lem}[thm]{Lemma}

\title{Difference equations for 
correlation functions of \\
Belavin's $\mathbb{Z}_n$-symmetric model
with boundary reflection}

\author{Yas-Hiro Quano}

\date{\it Department of Medical Electronics, 
Suzuka University of Medical Science \\
      \it Kishioka-cho, Suzuka 510-0293, Japan}
\begin{document}

\maketitle
\begin{abstract}
Belavin's $\mathbb{Z}_n$-symmetric 
elliptic model with boundary reflection is 
considered on the basis of the boundary CTM bootstrap. 
We find non-diagonal $K$-matrices 
for $n>2$ that satisfy the reflection equation 
(boundary Yang--Baxter equation), and also find 
non-diagonal Boltzmann weights for the 
$A^{(1)}_{n-1}$-face model even for $n\geqq 2$. 
We derive difference equations of 
the quantum Knizhnik-Zamolodchikov type for 
correlation functions of the boundary model. 
The boundary spontaneous polarization is obtained by solving 
the simplest difference equations. The resulting quantity 
is the square of the spontaneous polarization for the bulk 
$\mathbb{Z}_n$-symmetric model, up to a phase factor. 
\end{abstract}

\section{Introduction}
Integrable models with a boundary 
have been studied in massive quantum theories 
\cite{RE,GZ,SS,HSWY,dMM,FKQ} and half infinite 
lattice models \cite{Skl,JKKKM,JKKMW,MW,FK,KQ,H}. 
The boundary interaction is specified by the boundary 
$S$-matrix for massive quantum theories \cite{GZ}, 
and by the reflection matrix $K$ 
for lattice models \cite{Skl}. The integrability 
in the presence of reflecting boundary is ensured by 
the reflection equation (boundary Yang--Baxter 
equation) \cite{RE}, in addition to the Yang--Baxter 
equation for bulk (i.e., without boundary) theory \cite{ESM}. 

It was shown in \cite{GZ} that the boundary vacuum of 
boundary integrable theories can be expressed in terms of 
the vacuum and the creation operators in the bulk theory. 
In \cite{JKKKM} the explicit bosonic formulae of 
the boundary vacuum of the boundary XXZ model were 
obtained by using the bosonization of the vertex 
operators associated with the bulk 
XXZ model \cite{JM}. 

The quantum Knizhnik-Zamolodchikov equations 
\cite{Sm1,FR} are satisfied by both correlation functions
and form factors for bulk field theories \cite{S} and 
for bulk lattice models \cite{JMN,JKMQ} 
with the affine quantum group symmetry. 
It is shown in \cite{JKKMW} that correlation functions 
and form factors in semi-infinite XXZ/XYZ spin chains 
with integrable boundary conditions satisfy 
the boundary analogue of the quantum Knizhnik-Zamolodchikov 
equation \cite{RE}. 

In this paper we study Belavin's $\bz _n$-symmetric 
vertex model \cite{Bela} with integrable boundary 
condition, the boundary Belavin model. 
The $R$-matrix of Belavin's model is 
expressed in terms of elliptic functions of the 
spectral parameter $z$ so that the $R$-matrix has 
doubly quasi periodicity. Thus we expect that 
the $K$-matrix of the boundary Belavin model also 
possesses appropriate transformation properties with 
respect to $z$ compatible to those of the $R$-matrix. 
We shall show that under such assumption the $K$-matrix 
of the boundary Belavin model is inevitably non-diagonal 
for $n>2$. Our solution is diagonal for $n=2$ but 
different from the one used in \cite{JKKMW}. 

On the basis of boundary CTM bootstrap \cite{ESM,JMN,JKKMW} 
we find that the correlation functions for the boundary 
Belavin model satisfy a set of difference 
equations, the boundary analogue of the quantum 
Knizhnik--Zamolodchikov equation. Furthermore, 
by solving the simplest difference equations, 
we obtain the boundary spontaneous polarization 
which turns out to be the square of that for the 
bulk $\mathbb{Z}_n$-symmetric model \cite{SPn}. 

The rest of this paper is organized as follows.
In section 2 we review Belavin's 
$\mathbb{Z}_n$-symmetric model, 
thereby fixing our notations. In section 3 we give 
two non-diagonal solutions to the reflection equation, 
one is a constant $K$-matrix, and the other is an 
elliptic $K$-matrix. Furthermore, we consider the 
boundary analogue of the vertex-face correspondence to 
discuss the connection between our $K$-matrix and the 
boundary weights of the $A^{(1)}_{n-1}$ model \cite{BFKZ}. 
In section 4 we construct lattice realization of the 
boundary vacuum states and vertex operators from the 
boundary CTM bootstrap 
approach. In section 5 we derive difference equations for 
$N$-point functions of the boundary Belavin model. 
We solve the simplest difference equations with $N=1$ 
for free boundary condition to obtain the explicit 
expression of the boundary spontaneous polarization. 
The result gives the higher rank generalization of 
that for the boundary eight vertex model \cite{JKKMW}. 
In section 6 we summarize the results obtained 
in this paper, and give some concluding remarks. 

\section{Belavin's vertex model and the 
reflection equation}

\subsection{Elliptic theta functions}
For a complex number $\tau $ in the upper half-plane, 
let $\Lambda _{\tau }:=\bz +\bz \tau $ 
be the lattice generated by $1$ and $\tau$, 
and $E_{\tau }:= \bc /\Lambda _{\tau }$ the complex torus 
which can be identified with an elliptic curve. 
For $a, b\in \mathbb{R}$, introduce the Jacobi 
theta function
\begin{equation}
\vartheta\left[\begin{array}{c} a \\ b 
\end{array} \right](z,\tau ): 
= \displaystyle\sum_{m\in \bz } 
\exp \left\{ \pi \sqrt{-1}(m+a)~\left[ (m+a)\tau 
+2(z+b) \right] \right\}. \label{Rieth}
\end{equation}
Hereafter a positive integer $n \geq 2$ is fixed and 
we will use the following compact symbols
\begin{equation}
\sigma^{(n)}_{\mbox{\footnotesize\boldmath $\alpha $}}(z)=
\vartheta\left[\begin{array}{c} \alpha _2 /n +1/2 \\ 
\alpha _1 /n+1/2 
\end{array} \right](z,\tau ), ~~~~~~
\theta^{(j)}_n (z)=
\vartheta\left[\begin{array}{c} 1/2 - j/n \\ 1/2 \end{array}
\right](z,n\tau ),
\end{equation}
for $\mbox{\boldmath $\alpha $}=(\alpha _1,\alpha _2)\in 
\bz \otimes \bz$ and for $j\in \bz _n $; and 
$$
h(z):=\prod_{j=0}^{n-1} \theta^{(j)}(z)/
\prod_{j=1}^{n-1} \theta^{(j)}(0). 
$$
The superscript $(n)$ and the subscript $n$ 
will be often suppressed when we have no fear 
of confusion. 

The elliptic theta functions are expressed 
in terms of the product series 
\begin{equation}
\begin{array}{c}
\theta^{(j)}(z)=
\sqrt{-1}\omega ^{j/2}t^{n(1/2-j/n)^{2}} 
u^{-1+2j/n} 
(t^{2n}; t^{2n})_{\infty}
(t^{2j}u^{2}; t^{2n})_{\infty}
(t^{2(n-j)}u^{-2}; t^{2n})_{\infty}, \\
h(z)=t^{(n-1)/4}
\displaystyle\frac{(t^{2n}; t^{2n})^{3}_{\infty}}
{(t^{2}; t^{2})^{3}_{\infty}} 
\sigma_{\mbox{\footnotesize\boldmath $0$}}
(z,\tau )=\sqrt{-1}t^{n/4} 
\displaystyle\frac{(t^{2n}; t^{2n})^{3}_{\infty}}
{(t^{2}; t^{2})^{2}_{\infty}} 
u^{-1}
(u^{2}; t^{2})_{\infty}
(t^{2}u^{-2}; t^{2})_{\infty}, 
\end{array}\label{fpro}
\end{equation}
where 
$$
(a; q_{1}, \cdots , q_{k} )_{\infty } := 
\prod_{m_{1}=0}^{\infty} \cdots 
\prod_{m_{k}=0}^{\infty} (1-aq_{1}^{m_{1}}\cdots q_{k}^{m_{k}}). 
$$

\subsection{Belavin's vertex model}
Let $V=\bc ^n $ and $\{v_i \}_{i\in \bz _n }$ be 
the standard orthonormal basis of $V$ with the 
inner product $(v_j , v_k )=\delta_{jk}$. 
Let $V_z $ be a copy of $V$ with 
a spectral parameter $z$. 
The $\bz _n $-Baxter model is 
a vertex model on a two-dimensional square lattice ${\cal L}$ 
such that the state variables take on values of $\bz _n $-spin. 
Each oriented line of ${\cal L}$ 
carries a spectral parameter varying from line to line. 
We assign a $\bz _n $-valued local state on each edge. 
Let 

\begin{minipage}{\minitwocolumn}
\unitlength 1mm
\begin{picture}(40,30)
\put(47,15){$R(z_1 -z_2 )^{ik}_{jl}:=  
$}
\end{picture}
\end{minipage}%
\hspace{\columnsep}%
\begin{minipage}{\minitwocolumn}
\unitlength 1mm
\begin{picture}(40,30)
\put(-5,6){\begin{picture}(101,0)
\put(0,10){\vector(1,0){15}}
\put(15,10){\line(1,0){5}}
\put(10,0){\vector(0,1){15}}
\put(10,15){\line(0,1){5}}
\put(-3,9.5){$j$}
\put(9,21){$k$}
\put(9,-4){$l$}
\put(21,9.5){$i$}
\put(13,6){$z_{1}$}
\put(11,14){$z_{2}$}
\end{picture}
}
\end{picture}
\end{minipage}
be a local Boltzmann weight 
for a single vertex with bond states 
$i, j, k, l \in \bz _n$. 
Arrows denotes orientations of lines. 
We now define the linear map on $V_{z_1 }\otimes 
V_{z_2 }$ called the $R$-matrix as follows: 
\begin{eqnarray*}
R^{V_{z_1 }, V_{z_2 }}(v_j \otimes v_ l )=
\sum_{i, k \in \bz _n } 
(v_i \otimes v_k ) R(z_1 -z_2 )^{ik}_{jl}. 
\end{eqnarray*}

Belavin \cite{Bela} considered 
the $\bz _n $-symmetric model satisfying 
\begin{equation}
\begin{array}{cl}
\mbox{({\romannumeral 1})} & R(z)^{ik}_{jl}=0, 
\mbox{~~unless $i+k=j+l$,~~mod $n$}, \\
\mbox{({\romannumeral 2})} & R(z)^{i+p k+p}_{j+p l+p}=
R(z)^{ik}_{jl}, 
\mbox{~~for every $i,j,k,l$ and $p\in \bz _n$}.
\end{array} \label{Znsym}
\end{equation}
In terms of two linear map in $V$ 
\begin{equation}
gv_i =\omega ^i v_i ,~~~~~~
hv_i =v_{i-1}, \label{gh}
\end{equation}
where $\omega =\exp (2\pi \sqrt{-1}/n)$, the 
conditions (\ref{Znsym}) can be rephrased as follows: 
\begin{equation}
\begin{array}{rcl}
R(z)(g\otimes g) & = & (g\otimes g)R(z), \\
R(z)(h\otimes h) & = & (h\otimes h)R(z). 
\end{array} \label{gln}
\end{equation}
Thus the $R$-matrix of Belavin's 
$\bz _n $-symmetric model is of the form 
\begin{equation}
R(z)=\frac{1}{\kappa (z)}\overline{R}(z), ~~~~
\overline{R}(z)=
\sum_{\mbox{\footnotesize\boldmath $\alpha $}\in G_n} 
u_{\mbox{\footnotesize\boldmath $\alpha $}}(z)
I_{\mbox{\footnotesize\boldmath $\alpha $}} 
\otimes I_{\mbox{\footnotesize\boldmath $\alpha $}}^{-1}. 
\label{eq:Bel-sol}
\end{equation}
Here $G_n =\bz _n \otimes \bz_n$, and 
$I_{\mbox{\footnotesize\boldmath $\alpha $}}=
g^{\alpha_1}h^{\alpha_2}$ for 
$\mbox{\boldmath $\alpha $}=
(\alpha_1 , \alpha_2 )$. The normalization factor 
$\kappa (z)$ will be given lator. 
The coefficient function 
$u_{\mbox{\footnotesize\boldmath $\alpha $}}(z)$ 
is determined by imposing 
the $R$-matrix satisfies the Yang-Baxter equation 
\begin{eqnarray}
R_{12}(z_1 -z_2 )
R_{13}(z_1 -z_3 )
R_{23}(z_2 -z_3 )=
R_{23}(z_2 -z_3 )
R_{13}(z_1 -z_3 )
R_{12}(z_1 -z_2 ), \label{YBE}
\end{eqnarray}
where $R_{ij}(z)$ denotes the matrix on $V^{\otimes 3}$, 
which acts as $R(z)$ on the $i$-th and $j$-th components and 
as identity on the other one. 
Belavin's solution to (\ref{YBE}) is given as follows: 
\begin{equation}
u_{\mbox{\footnotesize\boldmath $\alpha $}}(z)
=u^{(n)}_{\mbox{\footnotesize\boldmath $\alpha $}}(z,w)
:=\frac{1}{n}
\frac{\sigma_{\mbox{\footnotesize\boldmath $\alpha $}}(z+w/n)}
{\sigma_{\mbox{\footnotesize\boldmath $\alpha $}}(w/n)}, 
\label{eq:u}
\end{equation}
where $w(\neq 0~\mbox{mod~}\Lambda _{\tau })$ is a constant. 
It is obvious that the following initial 
condition holds: 
\begin{equation}
\overline{R}(0)=P, ~~~~ 
P(x\otimes y)=y\otimes x. 
\label{eq:ini}
\end{equation}

In order to facilitate the derivation of the similar 
results for the $K$-matrix of the boundary 
$\bz _n$-symmetric model, we give brief sketches of 
proofs of several well known properties for 
Belavin's $R$-matrix. 
\begin{prop}~~~~~
The Boltzmann weights or the elements of 
$R$-matrix are given as follows \cite{RT}: 
\begin{equation}
\overline{R}(z)^{ik}_{jl}=\left\{ \begin{array}{ll} 
\displaystyle\frac{h(z)\theta ^{(i-k)}(z+w)}
{\theta ^{(j-k)}(z)\theta ^{(i-j)}(w)} & 
\mbox{if $i+k=j+l$, mod $n$,} \\
0 & \mbox{otherwise.} \end{array} \right. \label{Bel}
\end{equation}
\label{prop:Bel}
\end{prop}

[Proof] Because of the $\bz _n$-symmetry, 
\begin{eqnarray*}
R^{i-j k-j}_{0 l-j}(z)= 
R^{ik}_{jl}(z)=
\sum_{\mbox{\footnotesize\boldmath $\alpha $}\in G_n} 
u_{\mbox{\footnotesize\boldmath $\alpha $}}(z)
(I_{\mbox{\footnotesize\boldmath $\alpha $}})^{i}_{j} 
(I_{\mbox{\footnotesize\boldmath $\alpha $}}^{-1})^{k}_{l}=
\delta ^{i+k}_{j+l}\sum_{\alpha _1 \in \bz _n } 
u_{(\alpha _1 , j-i)}(z)\omega ^{(i-l)\alpha _1}.  
\end{eqnarray*}
Set ${\cal R}^{ab}(z)=R^{ab}_{0 a+b}(z)$. Then we have 
\begin{equation}
{\cal R}_n^{ab}(z,v)=\sum_{\alpha _1 \in \bz _n } 
u^{(n)}_{(\alpha _1 , -a)}(z,v)\omega ^{-b\alpha _1}. 
\label{eq:cal-R}
\end{equation}

The transformation property of ${\cal R}^{ab}(z)$ 
and the initial condition ${\cal R}^{ab}(0)
=\delta^{b0}$ imply that 
\begin{equation}
{\cal R}^{ab}(z)=0, ~~~~
\mbox{at $z=c\tau$ ($c\neq -b$, mod $n$) and 
$z=(a-b)\tau -w$, mod $\Lambda_{n\tau}$}. \label{R0}
\end{equation}
Hence ${\cal R}^{ab}(z)$ has the form
\begin{eqnarray*}
{\cal R}^{ab}(z)=C^{ab}(w)\theta ^{(a-b)}(z+w) 
\prod _{c\neq -b} \theta ^{(c)}(z).  
\end{eqnarray*}
By substituting $z=-b\tau $ we have 
\begin{eqnarray*}
C^{ab}(w)^{-1}=\theta ^{(a)}(w) 
\prod _{c\neq 0} \theta ^{(c)}(0),  
\end{eqnarray*}
which concludes that (\ref{Bel}) holds. $\Box$

As a corollary of Proposition \ref{prop:Bel} we have 
\cite{RT}
\begin{equation}
PR(-w)=-R(-w), ~~~~ 
R(w)P=R(w). \label{eq:PR}
\end{equation}

Now we assume that $0<t<q<u<1$, where 
$t:=\exp (\pi\sqrt{-1}\tau ), 
 q:=\exp (\pi\sqrt{-1}w)$, and $u:=\exp (-\pi\sqrt{-1}z)$. 
Following Baxter \cite{ESM} 
we call such domain of parameters the principal regime. 
Note that (\ref{Bel}) is weights of the eight-vertex 
model when $n=2$. 

\subsection{Unitarity and crossing symmetry}

Belavin's $R$-matrix satisfies the unitarity 
and crossing symmetry relations \cite{RT,Has,Th}. 

\begin{prop}~~~~~~Belavin's $R$-matrix satisfies 
the following unitarity relation or the first inversion 
relation: 
\begin{equation}
\overline{R}_{21}(z)\overline{R}_{12}(-z)
=\rho_1 (z,w)I\otimes I, \label{1st}
\end{equation}
where 
\begin{equation}
\rho_1 (z,w)=\frac{\sigma (z+w)\sigma (-z+w)}{\sigma ^2 (w)}. 
\label{uni}
\end{equation}
\label{5.3}
\end{prop}
[Proof] Note that 
\begin{eqnarray*}
\overline{R}_{21}(z)\overline{R}_{12}(-z) & = & 
\sum_{\mbox{\footnotesize\boldmath $\alpha $}\in G_n} 
u^{(n)}_{\mbox{\footnotesize\boldmath $\alpha $}}(z,w)
I_{\mbox{\footnotesize\boldmath $\alpha $}}^{-1} 
\otimes I_{\mbox{\footnotesize\boldmath $\alpha $}} 
\sum_{\mbox{\footnotesize\boldmath $\beta $}\in G_n} 
u^{(n)}_{\mbox{\footnotesize\boldmath $\beta $}}(-z,w)
I_{\mbox{\footnotesize\boldmath $\beta $}} 
\otimes I_{\mbox{\footnotesize\boldmath $\beta $}}^{-1} \\
& = & 
\sum_{\mbox{\footnotesize\boldmath $\alpha $}
      \mbox{\footnotesize\boldmath $\beta $}\in G_n } 
u^{(n)}_{\mbox{\footnotesize\boldmath $\alpha $}}(z)
u^{(n)}_{\mbox{\footnotesize\boldmath $\beta $}}(-z)
I_{\mbox{\footnotesize\boldmath $\alpha $}}^{-1} 
I_{\mbox{\footnotesize\boldmath $\beta $}} 
\otimes 
I_{\mbox{\footnotesize\boldmath $\alpha $}} 
I_{\mbox{\footnotesize\boldmath $\beta $}}^{-1} \\
& = & 
\sum_{\mbox{\footnotesize\boldmath $a$}\in G_n }
f^{(n)}_{\mbox{\footnotesize\boldmath $a$}}(z,w) 
I_{\mbox{\footnotesize\boldmath $a$}} 
\otimes 
I_{\mbox{\footnotesize\boldmath $a$}}^{-1}, 
\end{eqnarray*}
where 
\begin{equation}
f^{(n)}_{\mbox{\footnotesize\boldmath $a$}}(z,w) 
=\sum_{\mbox{\footnotesize\boldmath $\alpha $}\in G_n} 
\omega ^{\langle \mbox{\footnotesize\boldmath $\alpha $},
\mbox{\footnotesize\boldmath $a$}\rangle } 
u^{(n)}_{\mbox{\footnotesize\boldmath $\alpha $}}(z,w)
u^{(n)}_{\mbox{\footnotesize\boldmath $a$}+
\mbox{\footnotesize\boldmath $\alpha $}}(-z,w), 
\label{eq:f_a}
\end{equation}
and $\langle \mbox{\footnotesize\boldmath $\alpha $},
\mbox{\footnotesize\boldmath $a$}\rangle =
\alpha_1 a_2 -\alpha_2 a_1$. 
Proposition \ref{5.3} is thus reduced to 
\begin{equation}
f^{(n)}_{\mbox{\footnotesize\boldmath $a$}}(z,w) =
\rho_1 (z,w)\delta _{\mbox{\footnotesize\boldmath $a0$}}.
\label{eq:inv-1} 
\end{equation}
Concerning the proof of (\ref{eq:inv-1}), 
see Theorem 3.3 and Lemma 3.2 in \cite{Th}. $\Box$

~

Next we describe the crossing symmetry for Belavin's 
$\bz _n$-symmetric model. For that purpose 
let us recall the $R$-matrix on $K \otimes L$, where 
$K=V_{z_1 }\otimes \cdots \otimes V_{z_k }$ and 
$L=V_{z'_1 }\otimes \cdots \otimes V_{z'_l }$: 
$$
\begin{array}{rl}
R^{K, V_{z'}}:= & 
R^{V_{z_1 }, V_{z'}}_{1;k+1} 
\cdots 
R^{V_{z_k }, V_{z'}}_{k;k+1}, \\
R^{K,L}:= & 
R^{K, V_{z'_l }}_{1\cdots k; k+l} 
\cdots 
R^{K, V_{z'_1 }}_{1\cdots k; k+1}. 
\end{array} 
$$
YBE holds for $R^{K,L}$ 
by virtue of YBE for $R^{V,V}$ (\ref{YBE}) 
\begin{eqnarray}
R_{12}^{K, L}R_{13}^{K, M}R_{23}^{L, M}=
R_{23}^{L, M}R_{13}^{K, M}R_{12}^{K, L}, \label{KYBE}
\end{eqnarray}
as a linear map on $K\otimes L \otimes M$. 

For special $K^k_z=V_{z_1 }\otimes \cdots \otimes V_{z_k }$ 
such that $z_j =z+(k+1-j)w$ ($1\leqq j\leqq k$), 
the fusion operator $\pi$ associated with $K^k_z$ is given 
as follows \cite{fuse}: 
\begin{equation}
\pi :=R^{V_{z_1 }, V_{z_2}}_{k-1;k} 
R^{V_{z_1 }\otimes V_{z_2}, V_{z_3}}_{k-2,k-1;k} 
\cdots 
R^{V_{z_1 }\otimes \cdots \otimes V_{z_{k-1}}, 
V_{z_k }}_{1,\cdots ,k-1;k}. 
\label{eq:fuse}
\end{equation}
{}From the first equation of (\ref{eq:PR}) and 
the Yang--Baxter equation (\ref{YBE}) we have 
\begin{equation}
\pi (K^k_z )=\Lambda ^{k}(V)=\mbox{Anti}(K_z^k ). 
\end{equation}

Let $V^* $ be the dual space of $V$ and 
$\{v^{*}_i \}_{i\in \bz _n }$ be 
the dual basis of $\{v_i \}_{i\in \bz _n }$. 
Then we have the isomorphism 
$C: V^*_{z+nw/2} \longrightarrow 
\mbox{Anti}(K_z^{n-1})$
\begin{eqnarray}
Cv^* _i =\displaystyle\sum_{i_1 , \cdots , i_{n-1}}
\frac{\epsilon _i ^{i_1 \cdots i_{n-1}}}{\sqrt{(n-1)!}}
                 v_{i_1 }\otimes 
                 \cdots \otimes 
                 v_{i_{n-1}}, 
\label{hom}
\end{eqnarray}
where $\epsilon _i ^{i_1 \cdots i_{n-1}}$ is 
the $n$-th order completely antisymmetric tensor. 
The spectral parameter $z\!\!+\!\!nw/2$ associated with 
the dual space $V^*$ refers to the mean value 
of $n\!-\!1$ spectral parameters 
$z\!+\!(n-1)w$, $\cdots$, $z\!+\!w$ of 
$V$\footnote{Note that the spectral parameter of $V^*$
is shifted by $nw/2$ from the one in \cite{SPn,Th}. 
}. 
Then the $R$-matrices on $V\otimes V^*$ and 
$V^* \otimes V$ are defined as follows: 
\begin{equation}
\begin{array}{rcl}
R^{V_{z_1 }, V^{*}_{z_{2}+nw/2}} & = & 
(I\otimes C)^{-1}
R^{V_{z_1 }, 
V_{z_2 +(n-1)w}\otimes \cdots \otimes V_{z_2 +w}}
(I\otimes C), \\
R^{V^{*}_{z_1 +nw/2}, V_{z_{2}}} & = & 
(C\otimes I)^{-1}
R^{V_{z_1 +(n-1)w}\otimes \cdots 
\otimes V_{z_1 +w}, V_{z_2 }} 
(C\otimes I).
\end{array}
\label{VV*}
\end{equation}
The un-normalized $\overline{R}$ on $V\otimes V^*$ and 
$V^* \otimes V$ are also defined in a similar manner. 

\begin{prop}~~~~~The $R$-matrix on 
$V\otimes V^*$ and $V^* \otimes V$ defined 
in (\ref{VV*}) meet the crossing symmetry \cite{Has,Th}: 
\begin{equation}
\begin{array}{rcl}
\overline{R}_{21}^{V_{z_2 }, V^{*}_{z_{1}+nw/2}} 
& = & 
(\overline{R}_{12}^{V_{z_1 }, V_{z_2 }})^{t_1}  
\displaystyle\prod_{p=2}^{n-1} \frac{h(-z_1 +z_2 +pw)}{h(w)}, \\
\overline{R}_{12}^{V^{*}_{z_1 +nw/2}, 
V_{z_{2}}} & = & 
(\overline{R}_{21}^{V_{z_2 }, V_{z_1 +nw}})^{t_1}  
\displaystyle\prod_{p=1}^{n-2} \frac{h(-z_1 +z_2 -pw)}{h(w)}, 
\end{array} 
\label{ncross}
\end{equation}
where $t_i$ denotes the transposition of the $i$-th space. 
\label{prop:cross}
\end{prop}

[Proof] Let 
$$
\overline{R}_{21}^{V_{z_2}, V^{*}_{z_{1}+nw/2}}
(v_j \otimes v^{*}_{l})
= \sum_{i,k}(v_{i}\otimes v^{*}_{k}) \,
a^{ik}_{jl} (-z_1 +z_2 )
$$
Because of the initial condition (\ref{eq:ini}) and 
the second equation of (\ref{eq:PR}), 
the element $a^{ik}_{jl}(-z)$ vanishes at 
$-z=pw$, where $p=2, \cdots , n-1$. Thus we have 
an entire function $b^{ik}_{jl}(-z)$ 
from $a^{ik}_{jl}(-z)$ devided by 
$h(-z-2w) \cdots f(-z-(n-1)w)$. 

The transformation property of $b^{ik}_{jl}(-z)$ are 
the same as $\overline{R}^{li}_{kj}(z)$. It follows 
from the second equation of (\ref{eq:PR}) that 
$b^{ik}_{jl}(-z)=0$ at $z=c\tau$ for $c\neq j-k$ 
and at $z=(i-k)\tau -w$, which coincide the zeros 
of $\overline{R}^{li}_{kj}(z)$ (\ref{R0}). Thus 
$b^{ik}_{jl}(-z)$ equals $\check{R}^{kl}_{ij}(z)$ 
up to a scalar factor, which is determined by 
substituting $z=(k-i)\tau$. The second 
equation of (\ref{ncross}) can be shown in a 
similar way. $\Box$

{}From (\ref{uni}) and (\ref{ncross}), 
we have the following second inversion relation 
\cite{RT,Th} 
\begin{equation}
\sum_{jl} 
\overline{R}^{t_1}_{12}(z)
\overline{R}^{t_1}_{21}(-z-nw)=\rho_2 (z,w) I, 
\label{2nd}
\end{equation}
where 
\begin{equation}
\rho_2 (z,w)=\frac{h(-z)h(z+nw)}{h^2 (w)}. 
\end{equation}

Imposing the unitarity and crossing symmetry 
condition with respect to the normalized 
$R$-matrix: 
\begin{equation}
R_{21}(z)R_{12}(-z)
=I\otimes I, \label{eq:uni}
\end{equation}
\begin{equation}
R_{21}^{V_{z_2 }, V^{*}_{z_{1}+nw/2}} 
= (R_{12}^{V_{z_1 }, V_{z_2 }})^{t_1}, ~~~~
R_{12}^{V^{*}_{z_1 +nw/2}, 
V_{z_{2}}} = 
(R_{21}^{V_{z_2 }, V_{z_1 +nw}})^{t_1}, 
\label{eq:cross}
\end{equation}
the normalization factor $\kappa (z)$ should obey 
the following functional equations: 
\begin{equation}
\begin{array}{rcl}
\kappa (z) \kappa (-z) & = & \rho_1 (z,w), \\
\kappa (z) \kappa (-z-nw) & = & \rho_2 (z,w). 
\end{array} 
\label{inv}
\end{equation}

Hereafter $\kappa (z)$ is often denoted by $\kappa (u)$ 
through the relation $u=\exp (-\pi \sqrt{-1}z)$.  
In the principal regime using (\ref{fpro})  
the following expression solves (\ref{inv}) \cite{RT} 
\begin{equation}
\kappa (u)=u^{-(n-2)/n} 
\frac{(u^2 ; t^2 )_{\infty}(t^2 u^{-2} ; t^2 )_{\infty}}
     {(q^2 ; t^2 )_{\infty}(t^2 q^{-2} ; t^2 )_{\infty}} 
\bar{\kappa }(u), 
\end{equation}
where 
$$
\bar{\kappa }(u)= 
\frac{(q^2 u^2 ; t^2 , q^{2n})_{\infty}
      (q^{2n} u^{-2} ; t^2 , q^{2n})_{\infty}
      (t^2 q^{-2} u^2 ; t^2 , q^{2n})_{\infty}
      (t^2 q^{2n} u^{-2} ; t^2 , q^{2n})_{\infty}}
     {(q^{2+2n} u^{-2} ; t^2 , q^{2n})_{\infty}
      (u^2 ; t^2 , q^{2n})_{\infty}
      (t^2 q^{-2+2n} u^{-2} ; t^2 , q^{2n})_{\infty}
      (t^2 u^2 ; t^2 , q^{2n})_{\infty}}. 
$$

{}From $\kappa (1) =1$ the initial condition 
for $R$ also holds: 
\begin{equation}
R(0)=P. \label{ini}
\end{equation}

\section{Boundary Belavin model}

\subsection{Reflection equation for 
the boundary Belavin model} 

In this section we consider the following reflection 
equation or the boundary Yang--Baxter equation: 
\begin{equation}
K_2 (z_2 )R_{21}(z_1 +z_2 )K_1 (z_1 )R_{12}(z_1 -z_2 )
=
R_{21}(z_1 -z_2 )K_1 (z_1 )R_{12}(z_1 +z_2 )K_2 (z_2 ). 
\label{eq:RE}
\end{equation}
The reflection equation (\ref{eq:RE}) is valid 
when $z_1 =z_2$ because $R(0)=P$. Furthermore, 
the following Lemma holds: 
\begin{lem} The reflection equation $(\ref{eq:RE})$ 
is valid when $(1)$ $z_1 =0;$ $(2)$ $z_1 =-z_2$ provided 
\begin{equation}
\begin{array}{ll}
\mbox{$(1)$ Boundary initial condition$:$} 
& K(0)=I; \\ 
\mbox{$(2)$ Boundary unitarity relation$:$} 
& K(z)K(-z)=I, 
\end{array}
\label{eq:K-uni}
\end{equation}
respectively. 
\label{lem:sp-RE}
\end{lem}

[Proof] It is evident from the unitarity 
(\ref{eq:uni}) and the initial condition 
(\ref{ini}) for $R$-matrix. $\Box$

~

Here we notice that Belavin's $R$-matrix 
have the following quasi-periodic properties 
\begin{equation}
\begin{array}{rcl}
\overline{R}(z+1) & = & -(g\otimes I)^{-1} 
\overline{R}(z) (g\otimes I)
=(I\otimes g) \overline{R}(z) (I\otimes g)^{-1}, \\
\overline{R}(z+\tau ) & = & -(h\otimes I)^{-1} 
\overline{R}(z) (h\otimes I)
\times \exp \left\{ 
-2\pi \sqrt{-1}\left( z+\frac{\tau}{2}+\frac{w}{n} 
\right)\right\} \\
&=&-(I\otimes h) \overline{R}(z) (I\otimes h)^{-1}
\times \exp \left\{ 
-2\pi \sqrt{-1}\left( z+\frac{\tau}{2}+\frac{w}{n} 
\right)\right\}. \label{eq:R-quasi}
\end{array} 
\end{equation}
Thus we have the following Proposition: 
\begin{prop} Let 
$$
K(z)=\frac{1}{\lambda (z)} \overline{K}(z), 
$$
where $\lambda (z)$ is a scalor function. 
Suppose (\ref{eq:K-uni}) and 
the following quasi transformation property: 
\begin{equation}
\begin{array}{rcl}
\overline{K}(z+1) & = & -g
\overline{K}(z)g, \\
\overline{K}(z+\tau ) & = & -h
\overline{K}(z)h \times 
\exp \left\{ 
-2\pi \sqrt{-1}\left( z+\frac{\tau}{2}+c 
\right)\right\}, 
\end{array}
\label{eq:K-quasi}
\end{equation}
where $c$ is a constant. Then 
$\overline{K}(z)$ solves (\ref{eq:RE}). 
\label{prop:K-quasi}
\end{prop}

[Proof] Let $F(z_1 , z_2 )$ stand for the difference 
of the LHS and the RHS of (\ref{eq:RE}). Then we have 
\begin{equation}
\begin{array}{rcl}
F(z_1 +1 ,z_2 ) & = & -(g\otimes I) F(z_1 ,z_2 )
(g\otimes I), \\
F(z_1 +\tau ,z_2 ) & = & -(h\otimes I) F(z_1 ,z_2 )
(h\otimes I)\times \exp (-2\pi \sqrt{-1}B), 
\end{array}
\label{eq:RE-quasi}
\end{equation}
where $B=3z_1 +3\tau /2+2w/n+c$. The second equation 
of (\ref{eq:RE-quasi}) implies that the $(ik,jl)$-th 
element of $F(z_1 , z_2 )$ satisfies 
\begin{equation}
F(z_1 +\tau ,z_2 )^{ik}_{jl}=
-F(z_1 +\tau ,z_2 )^{i+1 k}_{j-1 l}\times 
(-2\pi \sqrt{-1}B). 
\label{eq:el-quasi}
\end{equation}
Thus we find that 
$F(p\tau ,z_2 )^{ik}_{jl}\propto 
F(0, z_2 )^{i+pk}_{j-pl}=0$ for $0\leqq p \leqq n-1$
from Lemma \ref{lem:sp-RE}. Similarly, we have 
$F(z_2 +p\tau ,z_2 )^{ik}_{jl}=
F(-z_2 +p\tau ,z_2 )^{ik}_{jl}=0$ for 
$0\leqq p \leqq n-1$: 
\begin{equation}
F(p\tau ,z_2 )^{ik}_{jl}=
F(z_2 +p\tau ,z_2 )^{ik}_{jl}=
F(-z_2 +p\tau ,z_2 )^{ik}_{jl}=0, 
~~~~ (0\leqq p \leqq n-1). 
\label{eq:zeroRE}
\end{equation}

Assume that $F(z_1 +\tau ,z_2 )^{ik}_{jl}$ is 
not identically zero. From Richey-Tracy's lemma 
(see section 3 in \cite{RT} or Lemma 2.4 in \cite{Th}) 
we conclude that $F(z_1 +\tau ,z_2 )^{ik}_{jl}$ has 
$3n$ zeros in $E_{n\tau}$ whose sum is equal to 
$nc -2w -3n(n-1)\tau -(i+j)\tau$. The contradiction 
to (\ref{eq:zeroRE}) implies the claim of 
this Proposition. $\Box$

\subsection{Solutions of the reflection equation}

Under the assumption of the quasi periodicity 
(\ref{eq:K-quasi}) compatible to (\ref{eq:R-quasi}) 
we find that 
the $K(z)$ is not a diagonal matrix for $n>2$. 
When $n=2$ we can take $K(z)$ diagonal because 
of $g^{-1}=g$ and $h^{-1}=h$. The most general and 
non-diagonal solution for $n=2$ is given in 
\cite{IK,HSFY}. Other non-diagonal solutions for 
$D^{(2)}_{n}$-vertex model are given in \cite{MG}. 

In this paper 
we consider the following two solutions of (\ref{eq:RE}), 
which can be also found in \cite{FSHY}. 

\subsubsection{Constant $K$-matrix} 
\begin{prop}
Let 
\begin{equation}
{\cal K}_0 v_j =v_{n-j}, 
\label{eq:cK}
\end{equation}
where $v_n =v_0$. 
Then ${\cal K}_0$ solves (\ref{eq:RE}). 
\label{prop:K0}
\end{prop}
[Proof] It is easy to see $g{\cal K}_0 g
=h{\cal K}_0 h={\cal K}_0$. Hence we have 
$$
\begin{array}{cl}
&K_2 (z_2 )R_{21}(z_1 +z_2 )K_1 (z_1 )R_{12}(z_1 -z_2 ) \\
=&I\otimes {\cal K}_0 
\displaystyle\sum_{\mbox{\footnotesize\boldmath $\alpha $}}
u_{\mbox{\footnotesize\boldmath $\alpha $}}(z_1 +z_2 )
(I_{\mbox{\footnotesize\boldmath $\alpha $}}^{-1} 
\otimes I_{\mbox{\footnotesize\boldmath $\alpha $}}) 
({\cal K}_0 \otimes I) 
\sum_{\mbox{\footnotesize\boldmath $\beta $}}
u_{\mbox{\footnotesize\boldmath $\beta $}}(z_1 -z_2 )
(I_{\mbox{\footnotesize\boldmath $\beta $}} 
\otimes 
I_{\mbox{\footnotesize\boldmath $\beta $}}^{-1}) \\ 
=&{\cal K}_0 \otimes {\cal K}_0 
\displaystyle\sum_{\mbox{\footnotesize\boldmath $\alpha $}}
\omega^{\alpha_1 \alpha_2}
u_{\mbox{\footnotesize\boldmath $\alpha $}}(z_1 +z_2 )
I_{\mbox{\footnotesize\boldmath $\alpha $}} 
\otimes I_{\mbox{\footnotesize\boldmath $\alpha $}} 
\sum_{\mbox{\footnotesize\boldmath $\beta $}}
\omega^{\beta_1 \beta_2}
u_{\mbox{\footnotesize\boldmath $\beta $}}(z_1 -z_2 )
I_{\mbox{\footnotesize\boldmath $\beta $}} 
\otimes 
I_{-\mbox{\footnotesize\boldmath $\beta $}} \\ 
=&{\cal K}_0 \otimes {\cal K}_0 
\displaystyle\sum_{\mbox{\footnotesize\boldmath $\beta $}}
\omega^{\beta_1 \beta_2}
u_{\mbox{\footnotesize\boldmath $\beta $}}(z_1 -z_2 )
I_{\mbox{\footnotesize\boldmath $\beta $}} 
\otimes 
I_{-\mbox{\footnotesize\boldmath $\beta $}} 
\sum_{\mbox{\footnotesize\boldmath $\alpha $}}
\omega^{\alpha_1 \alpha_2}
u_{\mbox{\footnotesize\boldmath $\alpha $}}(z_1 +z_2 )
I_{\mbox{\footnotesize\boldmath $\alpha $}} 
\otimes I_{\mbox{\footnotesize\boldmath $\alpha $}} \\ 
=&
\displaystyle\sum_{\mbox{\footnotesize\boldmath $\beta $}}
\omega^{\beta_1 \beta_2}
u_{\mbox{\footnotesize\boldmath $\beta $}}(z_1 -z_2 )
(I_{-\mbox{\footnotesize\boldmath $\beta $}} 
\otimes 
I_{\mbox{\footnotesize\boldmath $\beta $}}) 
({\cal K}_0 \otimes I) 
\sum_{\mbox{\footnotesize\boldmath $\alpha $}}
\omega^{\alpha_1 \alpha_2}
u_{\mbox{\footnotesize\boldmath $\alpha $}}(z_1 +z_2 )
(I_{\mbox{\footnotesize\boldmath $\alpha $}} 
\otimes 
I_{-\mbox{\footnotesize\boldmath $\alpha $}})
(I \otimes {\cal K}_0 )\\
=&
R_{21}(z_1 -z_2 )K_1 (z_1 )R_{12}(z_1 +z_2 )K_2 (z_2 ), 
\end{array}
$$
that implies this Proposition. $\Box$ 

\subsubsection{Elliptic $K$-matrix}

Let 
$$
m=\left\{ \begin{array}{ll} 
n & \mbox{if $n$ is odd,} \\
n/2 & \mbox{if $n$ is even}, 
\end{array} \right. 
$$
and let 
\begin{equation}
{\cal K}(z)=\sum_{
\mbox{\footnotesize\boldmath $\alpha $}\in G_m} 
\omega^{2\alpha_1 \alpha_2} 
u^{(n)}_{2\mbox{\footnotesize\boldmath $\alpha $}}
(z,v) I_{2\mbox{\footnotesize\boldmath $\alpha $}}
=\sum_{
\mbox{\footnotesize\boldmath $\alpha $}\in G_m} 
u^{(n)}_{2\mbox{\footnotesize\boldmath $\alpha $}}
(z,v) J_{\mbox{\footnotesize\boldmath $\alpha $}}, 
\end{equation}
where 
$$
J_{\mbox{\footnotesize\boldmath $\alpha $}}
=h^{\alpha_2}g^{2\alpha_1}h^{\alpha_2}
$$
for 
$\mbox{\boldmath$\alpha$}=
(\alpha_1 , \alpha_2 )$, and 
$v(\neq 0~\mbox{mod~}\Lambda _{\tau })$ is a constant. 
Using the identity 
$$
\frac{1}{m}\sum_{\alpha_1 =0}^{m-1} \omega^{2\alpha_1 
(i-\alpha_2 )} =\left\{ \begin{array}{ll}
\delta_{\alpha_2 , i} & \mbox{if $n$ is odd,} \\
\delta_{\alpha_2 , i}+\delta_{\alpha_2 , i-m} 
& \mbox{if $n$ is even,}
\end{array} \right. 
$$
we have ${\cal K}(0)={\cal K}_0$. 

\begin{lem} The following quasi transformation 
property holds: 
\begin{equation}
\begin{array}{rcl}
{\cal K}(z+1) &=&-g^{-1}{\cal K}(z)g; \\
{\cal K}(z+\tau ) &=&-h^{-1}{\cal K}(z)h
\times \exp \left\{ -2\pi\sqrt{-1} 
\left( z+\frac{\tau}{2}+\frac{v}{m} \right) 
\right\}. 
\end{array}
\label{eq:cal-K-quasi}
\end{equation}
\label{lem:cal-K-quasi}
\end{lem}

[Proof] This is based on the transformation 
properties of the elliptic theta function. 
$\Box$

\begin{lem} Let $\overline{K}(z)={\cal K}_0
{\cal K}(z)$. Then the boundary inversion relation 
holds: 
\begin{equation}
\overline{K}(z)\overline{K}(-z)=
\rho_1 (z,v) I. 
\label{eq:K-inv}
\end{equation}
\label{lem:K-uni}
\end{lem}
[Proof] Direct calculation shows 
$$
\begin{array}{rcl}
\overline{K}(z)\overline{K}(-z)&=&
{\cal K}_0 \displaystyle\sum_{
\mbox{\footnotesize\boldmath $\alpha $}\in G_m} 
u^{(n)}_{2\mbox{\footnotesize\boldmath $\alpha $}}
(z,v) J_{\mbox{\footnotesize\boldmath $\alpha $}} 
{\cal K}_0 
\displaystyle\sum_{
\mbox{\footnotesize\boldmath $\beta $}\in G_m} 
u^{(n)}_{2\mbox{\footnotesize\boldmath $\beta $}}
(-z,v) J_{\mbox{\footnotesize\boldmath $\beta $}} \\ 
&=& \displaystyle\sum_{
\mbox{\footnotesize\boldmath $\alpha $}\in G_m} 
u^{(n)}_{2\mbox{\footnotesize\boldmath $\alpha $}}
(z,v) J_{-\mbox{\footnotesize\boldmath $\alpha $}} 
\displaystyle\sum_{
\mbox{\footnotesize\boldmath $\beta $}\in G_m} 
u^{(n)}_{2\mbox{\footnotesize\boldmath $\beta $}}
(-z,v) J_{\mbox{\footnotesize\boldmath $\beta $}} \\ 
&=& \displaystyle\sum_{
\mbox{\footnotesize\boldmath $\alpha $}\in G_m} 
\sum_{
\mbox{\footnotesize\boldmath $\beta $}\in G_m} 
\omega^{2\langle 
\mbox{\footnotesize\boldmath $\alpha $}, 
\mbox{\footnotesize\boldmath $\beta $} \rangle}
u^{(n)}_{\mbox{2\footnotesize\boldmath $\alpha $}}
(z,v) u^{(n)}_{2\mbox{\footnotesize\boldmath $\beta $}}
(-z,v) J_{\mbox{\footnotesize\boldmath $\alpha $}
-\mbox{\footnotesize\boldmath $\beta $}} \\
& = & \displaystyle\sum_{
\mbox{\footnotesize\boldmath $a$}\in G_m }
g^{(n)}_{\mbox{\footnotesize\boldmath $a$}}(z,v) 
I_{\mbox{\footnotesize\boldmath $a$}}, 
\end{array}
$$
where 
\begin{equation}
g^{(n)}_{\mbox{\footnotesize\boldmath $a$}}(z,v) 
=\sum_{\mbox{\footnotesize\boldmath $\alpha $}\in G_m} 
\omega ^{2\langle \mbox{\footnotesize\boldmath $\alpha $},
\mbox{\footnotesize\boldmath $a$}\rangle } 
u^{(n)}_{2\mbox{\footnotesize\boldmath $\alpha $}}(z,v)
u^{(n)}_{2(\mbox{\footnotesize\boldmath $a$}+
\mbox{\footnotesize\boldmath $\alpha $})}(-z,v). 
\end{equation}
By comparing 
$g^{(n)}_{\mbox{\footnotesize\boldmath $a$}}(z,v)$ 
with 
$f^{(n)}_{\mbox{\footnotesize\boldmath $a$}}(z,w)$ 
defined in (\ref{eq:f_a}), we easily have 
$g^{(n)}_{\mbox{\footnotesize\boldmath $a$}}(z,v)
=f^{(m)}_{\mbox{\footnotesize\boldmath $a$}}(z,v)$ 
and hence (\ref{eq:K-inv}) holds for even $n$. 
Repeating the similar argument in Proposition \ref{5.3} 
we can also obtain (\ref{eq:K-inv}) 
for odd $n$. $\Box$ 

\begin{thm} Let $\overline{K}(z)={\cal K}_0
{\cal K}(z)$. Then $\overline{K}(z)$ solves 
the reflection equation (\ref{eq:RE}). 
\label{thm:ellip-K}
\end{thm}
[Proof] From Lemma \ref{lem:cal-K-quasi} 
we find that $\overline{K}(z)$ satisfies 
(\ref{eq:K-quasi}) with $c=v/m$. Since 
$\overline{K}(0)={\cal K}_0{}^2=I$, the 
$\overline{K}(z)$ also satisfies the first 
equation of (\ref{eq:K-uni}). It follows from 
Lemma \ref{lem:K-uni} that $\overline{K}(z)$ 
satisfies the second one of (\ref{eq:K-uni}). 
Thus $\overline{K}(z)$ is a solution to 
the reflection equation (\ref{eq:RE}) from 
Proposition \ref{prop:K-quasi}. $\Box$ 

\noindent{\bf Remark.} Our $K$-matrix for $n=2$ 
is different from the one used in \cite{JKKMW} 
so that the readers should be careful to compare 
our results with those of $n=2$. 

\subsection{Matrix elements of $K$-matrix}

In this subsection we calculate the $(j,k)$-th element 
of $\overline{K}(z)$: 
$$
\overline{K}(z)v_k =\sum_{j\in \bz_n} 
v_j \overline{K}(z)^j_k . 
$$
Note that 
$$
\overline{K}(z)^j_k = 
{\cal K}(z)^{n-j}_{k} =
\displaystyle\sum_{\alpha_2 \in \bz_m} 
\delta^{2\alpha_2}_{j+k} 
\displaystyle\sum_{\alpha_1 \in \bz_m} 
u^{(n)}_{(2\alpha_1 , j+k)}
(z,v) \omega^{-(j-k)\alpha_1} 
$$
When $n$ is even, thanks to the sum over $\alpha_2$, 
$\overline{K}(z)^j_k =0$ if $j+k$ is odd. 
By comparing (\ref{eq:cal-R}) we obtain 
\begin{equation}
\overline{K}(z)^j_k =
\left\{ \begin{array}{ll} 
{\cal R}_m^{-\frac{j+k}{2}, \frac{j-k}{2}}(z,v) 
& \mbox{if $j+k$ is even}, \\
0 & \mbox{if $j+k$ is odd}, 
\end{array} \right. 
\label{eq:K-even}
\end{equation}
for even $n$, 
\begin{equation}
\overline{K}(z)^j_k =
\left\{ \begin{array}{ll} 
{\cal R}_n^{-j-k, \frac{j-k}{2}}(z,v) 
& \mbox{$j-k$ is even}, \\
{\cal R}_n^{-j-k, \frac{j-k+n}{2}}(z,v) 
& \mbox{$j-k$ is odd}, 
\end{array} \right. 
\label{eq:K-odd}
\end{equation}
for odd $n$. 

We are now in a position to determine the normalization 
factor $\lambda (z)$. The boundary inversion relation 
(\ref{eq:K-inv}) implies 
\begin{equation}
\lambda (z)\lambda (-z)=\rho_1 (z,v). 
\label{eq:l-inv}
\end{equation}
Furthermore, the boundary crossing symmetry 
holds for $n=2$ \cite{Skl,GZ,JKKKM,JKKMW}: 
\begin{equation}
K (z)^j_k =\sum_{j',k'} 
R(-2z-w)^{j'\,1-k'}_{1-j\,k}K(-z-w)_{j'}^{k'}, 
\label{eq:b-x}
\end{equation}
which implies that 
\begin{equation}
\frac{\lambda (-z-w)}{\lambda (z)}=
\frac{1}{\bar{\kappa}(u^2 )} 
\frac{(q^2 u^{-2}; t^2 )_\infty 
(t^2 q^{-2}u^2 ; t^2 )_\infty }
{(u^2 ; t^2 )_\infty (t^2 u^{-2}; t^2 )_\infty }. 
\label{eq:l-cross}
\end{equation} 
Since $V^* \cong \Lambda^{n-1}(V) \not{\!\!\!\:\cong} V$ 
for $n>2$, the LHS of (\ref{eq:b-x}) for higher $n$ 
should be replaced by the $(j,k)$-th element of 
the dual $K$-matrix. 
We wish to discuss this point again in section 4. 

Here we assume the following functional 
relation holds for $n\geqq 2$: 
\begin{equation}
\frac{\lambda (-z-\tfrac{n}{2}w)}{\lambda (z)}=
\frac{1}{\bar{\kappa}(u^2 )} 
\frac{(q^n u^{-2}; t^2 )_\infty 
(t^2 q^{-n}u^2 ; t^2 )_\infty }
{(u^2 ; t^2 )_\infty (t^2 u^{-2}; t^2 )_\infty }. 
\label{eq:ln-cross}
\end{equation} 
Not (\ref{eq:ln-cross}) but (\ref{eq:l-inv}) 
is important to calculate the spontaneous polarization 
in section 5, so that we proceed further under the 
assumption (\ref{eq:ln-cross}). 
By solving (\ref{eq:l-inv}) and (\ref{eq:l-cross}) 
we obtain 
\begin{equation}
\lambda (z) = 
\frac{1}{(r^2 ; t^2 )_\infty 
(t^2 r^{-2}; t^2 )_\infty} 
\frac{(r^{2}u^{2}; t^{2}, q^{2n})_\infty 
(t^{2}r^{-2}u^{2}; t^{2}, q^{2n})_\infty }
{(r^{2}q^{2n}u^{-2}; t^{2}, q^{2n})_\infty 
(t^{2}r^{-2}q^{2n}u^{-2}; t^{2}, q^{2n})_\infty }
\frac{\phi (u^2 )}{\phi (u^{-2})}, 
\label{eq:exp-l}
\end{equation}
where $r=\exp (-\pi \sqrt{-1}v)$, and 
$$
\begin{array}{rcl}
\phi (x)&=&\displaystyle\frac{
(q^{n}x; t^{2}, q^{2n})_\infty 
(t^{2}q^{n}x; t^{2}, q^{2n})_\infty }
{(q^{2n}x; t^{2}, q^{2n})_\infty 
(t^{2}x; t^{2}, q^{2n})_\infty 
(r^{2}q^{n}x; t^{2}, q^{2n})_\infty 
(t^{2}r^{-2}q^{n}x; t^{2}, q^{2n})_\infty } \\[3mm]
&\times&\displaystyle\frac{
(q^{2n+2}x^{2}; t^{2}, q^{4n})_\infty 
(t^{2}q^{2n-2}x^{2}; t^{2}, q^{4n})_\infty }
{(q^{2n}x^{2}; t^{2}, q^{4n})_\infty 
(t^{2}q^{2n}x^{2}; t^{2}, q^{4n})_\infty }. 
\end{array}
$$

\subsection{Comments on boundary weights for the 
boundary $A^{(1)}_{n-1}$ face model}

In this subsection we wish to discuss the boundary 
analogue of the vertex-face correspondence. Concerning 
the case $n=2$, see \cite{H,FHS}. 
Let us consider the bulk $A^{(1)}_{n-1}$-face model 
whose local state takes on values of $P$, the weight 
lattice of $A^{(1)}_{n-1}$ \cite{JMO}. 
An ordered pair $(a,b) \in P^2$ is called 
admissible if $b=a+\hat{j}$, 
for a certain $j\in \bz_n$, where 
$$
\hat{j} =v_j -\frac{1}{n}\sum_{k=0}^{n-1}v_k . 
$$
Let 

\unitlength 1mm
\begin{picture}(100,20)
\put(20,7){$W
\left( \left. \begin{array}{cc} a & b 
\\ d & c \end{array} 
\right| z_1 -z_2 \right)\,=$} 
\put(65,3){\begin{picture}(101,0)
\put(0,0){\line(1,0){10}}
\put(10,0){\line(0,1){10}}
\put(0,0){\line(0,1){10}}
\put(0,10){\line(1,0){10}}
\multiput(-2,5)(2.2,0){6}{\line(1,0){1.2}}
\put(11.2,5){\vector(1,0){1,2}}
\put(12.8,4.2){$z_1$}
\multiput(5,-2)(0,2.2){6}{\line(0,1){1.2}}
\put(5,11.2){\vector(0,1){1,2}}
\put(4,13.5){$z_2$}
\put(-2,10.5){$a$}
\put(10.5,10.1){$b$}
\put(10.5,-1.5){$c$}
\put(-2.3,-1.8){$d$}
\end{picture}
}
\end{picture}

\noindent be the local Boltzmann weight for a 
state configuration $(a,b,c,d)$ round a face. 
Then $W
\left( \left. \begin{array}{cc} 
a & b \\ d & c \end{array} \right| 
z \right) =0~~$ unless all the four pairs 
$(a,b), (a,d), (b,c)$ and $(d,c)$ are admissible. 
Non-zero Boltzmann weights are given as follows: 
\begin{equation}
W
\left( \left. \begin{array}{cc} 
a & b \\ d & c \end{array} \right| 
z \right)=\frac{1}{w(z,w)}\overline{W}
\left( \left. \begin{array}{cc} 
a & b \\ d & c \end{array} \right| 
z \right), 
\end{equation}
where $w(z,w)$ is a scalar function and 
\begin{equation}
\begin{array}{rcl}
\overline{W}
\left( \left. \begin{array}{cc} 
a & a+\hat{j} \\ 
a+\hat{j} & a+2\hat{j} \end{array}
\right| z  \right) 
& = & \dfrac{h(z+w)}{h(w)}, \\
~ & ~ & ~ \\
\overline{W}
\left( \left. \begin{array}{cc} 
a & a+\hat{j} \\
a+\hat{j} & a+\hat{j}+\hat{k} \end{array}
\right| z \right) & = & 
\dfrac{h(a_{jk}w-z)}{h(a_{jk}w)}
~~~~(j \neq k), \label{BW} \\
~ & ~ & ~ \\
\overline{W}
\left( \left. \begin{array}{cc} 
a & a+\hat{k} \\ 
a+\hat{j} & a+\hat{j}+\hat{k} \end{array}
\right|z \right) & = & 
\dfrac{h(z)}{h(w)}\dfrac{h(a_{jk}w+w)}{h(a_{jk}w)}
~~~~(j\neq k). 
\end{array}
\end{equation}
Here 
$$
a_{jk}=\bar{a_j}-\bar{a_k}, ~~~~
\bar{a_j}=(a+\rho , v_j ), 
$$
and $\rho =\displaystyle\sum_{j=0}^{n-1} 
(n-1-j) \hat{j}$ is the half sum of the positive roots. 

Jimbo, Miwa and Okado \cite{JMO} 
introduced the intertwining vectors 
to show the equivalence between 
the $\bz _n$-symmetric model and 
the $A^{(1)}_{n-1}$ model. Let 
\begin{equation}\begin{array}{rl}
t^{a}_{b}(z):=  & 
~^{t}(t_{b}^{a(0)}(z), \cdots , t_{b}^{a(n-1)}(z)), \\
t_{b}^{a(i)}(z):= & \left\{ \begin{array}{ll} 
\theta^{(i)}(z+\delta -nw\bar{a}_j ) ~~ & 
\mbox{if $b=a+\bar{\epsilon}_j ,$} \\
0 ~~ & \mbox{otherwise,} \end{array} \right. 
\end{array} \label{phi}
\end{equation}
where $\delta$ is an arbitrary constant. Then we have 
the so-called vertex-face correspondence \cite{JMO}: 
\begin{equation}
\overline{R}(z_1 -z_2 ) 
t^{a}_{d}(z_1 )\otimes t^{d}_{c}(z_2 )=
\sum_{b} \overline{W}\left( \left. \begin{array}{cc} 
a & b \\ d & c \end{array} \right| z_1 -z_2 \right) 
t^{b}_{c}(z_1 )\otimes t^{a}_{b}(z_2 ). 
\label{eq:JMO}
\end{equation}
Thanks to (\ref{eq:JMO}) the Boltzmann weights 
(\ref{BW}) solve the face-type Yang-Baxter 
equation \cite{JMO}: 
\begin{equation}
\begin{array}{cc}
~ & 
\displaystyle \sum_{g} 
\overline{W}\left( \left. 
\begin{array}{cc} b & c \\ g & d \end{array} 
\right| z_1 -z_2 \right)
\overline{W}\left( \left. \begin{array}{cc} 
a & b \\ f & g \end{array} \right| z_1 -z_3 \right) 
\overline{W}\left( \left. \begin{array}{cc} 
f & g \\ e & d \end{array} \right| z_2 -z_3 \right) \\
~ & ~ \\
= & 
\displaystyle \sum_{g} 
\overline{W}\left( \left. \begin{array}{cc} 
a & b \\ g & c \end{array} \right| z_2 -z_3 \right) 
\overline{W}\left( \left. \begin{array}{cc} 
g & c \\ e & d \end{array} \right| z_1 -z_3 \right) 
\overline{W}\left( \left. \begin{array}{cc} 
a & g \\ f & e \end{array} \right| z_1 -z_2 \right). 
\end{array} \label{eq:WYBE}
\end{equation}

~

Let us now consider the boundary $A^{(1)}_{n-1}$-face 
model. By analogy with the bulk case, we find 
the following Proposition: 
\begin{prop} Assume that the existence of 
boundary weights $V$'s satisfying 
\begin{equation}
\overline{K}(z) t^a_c (z) =\sum_b 
\overline{V}\left( \left. a \begin{array}{c} b \\ c 
\end{array} \right| z \right) t^a_b (-z). 
\label{eq:b-vf}
\end{equation}
Then $\overline{V}$ solves the face-type reflection equation 
\begin{equation}
\begin{array}{cl}
&\displaystyle\sum_{b,e} 
\overline{V}\left( \left. f \begin{array}{c} g \\ e 
\end{array} \right| z_2 \right)
\overline{W}\left( \left. \begin{array}{cc} 
a & f \\ b & e \end{array} \right| z_1 +z_2 \right)
\overline{V}\left( \left. b \begin{array}{c} e \\ c 
\end{array} \right| z_1 \right) 
\overline{W}\left( \left. \begin{array}{cc} 
a & b \\ d & c \end{array} \right| z_1 -z_2 \right) \\
\\
=& \displaystyle\sum_{b,e} 
\overline{W}\left( \left. \begin{array}{cc} 
a & f \\ b & g \end{array} \right| z_1 -z_2 \right)
\overline{V}\left( \left. b \begin{array}{c} g \\ e 
\end{array} \right| z_1 \right)
\overline{W}\left( \left. \begin{array}{cc} 
a & b \\ d & e \end{array} \right| z_1 +z_2 \right)
\overline{V}\left( \left. d \begin{array}{c} e \\ c 
\end{array} \right| z_2 \right). 
\end{array}
\label{eq:face-RE}
\end{equation}
\label{prop:b-vf}
\end{prop}

In order to solve (\ref{eq:b-vf}), let us recall 
the dual intertwining vectors \cite{QF,Th,Has} 
\begin{equation}
\begin{array}{rl}
t_{a}^{*b}(z):= & 
(t_{a(0)}^{*b}(z), \cdots , t_{a(n-1)}^{*b}(z)), \\
t^{*a+\hat{j}}_{a(i)}(z):= & 
(\tilde{\Phi }^{a}(z))_{j}^{i}/\det \Phi ^a (z). 
\end{array} \label{eq:t-bar}
\end{equation}
Here $\Phi ^a (z)$ is a matrix whose $(i,j)$-component 
is $t_{a+\hat{j}}^{a(i)}(z)$, and 
$\tilde{\Phi }^a (z)$ is a cofactor matrix of $\Phi ^a (z)$. 
Note that $t^{a}_{b}(z)$ is a column vector while 
$t_{a}^{*b}(z)$ is a row vector. 
Thus by the rule of multiplication of matrices, 
$t_{a}^{*b}(z)t^{c}_{d}(z')$ represents a scalar function 
while $t^{a}_{b}(z)t^{*c}_{d}(z')$ 
does a function-valued matrix. 
Since $t^{a}_{b}(z)$ and $t^{*b}_{a}(z)$ 
enjoy the following orthogonal properties 
\begin{eqnarray}
t_{a}^{*a+\hat{j}}(z)t^{a}_{a+\hat{k}}(z) 
& = & 
\delta_{jk}, \label{eq:t*t} \\
\sum_{j=0}^{n-1} t^{a}_{a+\bar{\epsilon }_j }(z)
t_{a}^{*a+\hat{j}}(z) 
& = & I_n , \label{eq:tt*}
\end{eqnarray}
the boundary analogue of the vertex-face correspondence 
(\ref{eq:b-vf}) is equivalent to 
\begin{equation}
\overline{V}\left( \left. a \begin{array}{c} b \\ c 
\end{array} \right| z \right)=
t^{*b}_{a}(-z)K(z)t^{a}_{c}(z)
=\sum_{j,k} t^{*b}_{a(j)}(-z)K(z)^j_k t^{a(k)}_{c}(z). 
\label{eq:V-a-bc}
\end{equation}

\begin{prop} Let 
$$
V\left( \left. a \begin{array}{c} b \\ c 
\end{array} \right| z \right)=\frac{1}{\lambda (z)}
\overline{V}\left( \left. a \begin{array}{c} b \\ c 
\end{array} \right| z \right), 
$$
where $\lambda (z)$ is the same scalar function as 
for $K(z)$, and $\overline{V}$ is defined by 
(\ref{eq:V-a-bc}). Then the boundary weights $V$'s 
satisfy the initial condition 
\begin{equation}
V\left( \left. a \begin{array}{c} b \\ c 
\end{array} \right| 0 \right)=\delta_c^b , 
\label{eq:V-ini}
\end{equation}
and the inversion relation 
\begin{equation}
\sum_g V\left( \left. a \begin{array}{c} b \\ g 
\end{array} \right| z \right)
V\left( \left. a \begin{array}{c} g \\ c 
\end{array} \right| -z \right)=\delta_c^b . 
\label{eq:V-inv}
\end{equation}
\label{prop:V-ini/v}
\end{prop}

[Proof] The initial condition (\ref{eq:V-ini}) 
follows from that for $K(z)$ and (\ref{eq:t*t}). 
The inversion relation (\ref{eq:V-inv}) follows 
from (\ref{eq:tt*}), (\ref{eq:K-inv}) and 
(\ref{eq:t*t}). $\Box$

The boundary weights 
$V\left( \left. a \begin{array}{c} b \\ c 
\end{array} \right| z \right)$ are non-diagonal 
in the sense that they do not vanish even 
for $b\neq c$ as a function of $z$. Hence (\ref{eq:V-a-bc}) 
does not coincide with the 
diagonal solution of (\ref{eq:face-RE}) involving the 
bulk Boltzmann weights for the $A^{(1)}_{n-1}$-face 
model given in \cite{BFKZ} for $n\geqq 2$. 
Such disagreement indicates 
that there may exist unknown solution to (\ref{eq:RE}) 
corresponding to the solution given in \cite{BFKZ} 
and also unknown solution to (\ref{eq:face-RE}) 
corresponding to our $K$-matrix, throughout the 
boundary vertex-face correspondence. 

\subsection{Commuting transfer matrix}

The transfer matrix with $L$ columns, 

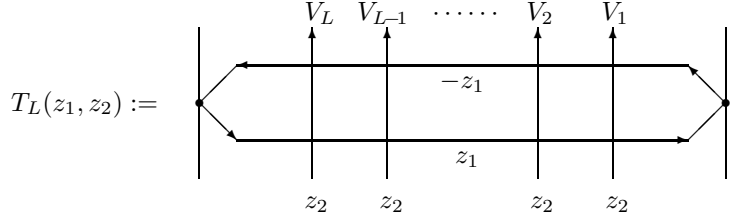
\unitlength 1mm
\begin{picture}(80,35)
\put(20,14){$T_L (z_1 , z_2 ):=$}
\put(40,0){\begin{picture}(101,0)
\put(10,20){\line(-1,-1){5}}
\put(5,15){\vector(1,-1){5}}
\put(10,10){\vector(1,0){60}}
\put(70,10){\line(1,1){5}}
\put(75,15){\vector(-1,1){5}}
\put(70,20){\vector(-1,0){60}}
\put(20,5){\vector(0,1){20}}
\put(30,5){\vector(0,1){20}}
\put(50,5){\vector(0,1){20}}
\put(60,5){\vector(0,1){20}}
\put(5,5){\line(0,1){20}}
\put(5,15){\circle*{1}}
\put(75,5){\line(0,1){20}}
\put(75,15){\circle*{1}}
\put(39,7){$z_{1}$}
\put(37,17){$-z_{1}$}
\put(19,1){$z_{2}$}
\put(29,1){$z_{2}$}
\put(49,1){$z_{2}$}
\put(59,1){$z_{2}$}
\put(19,26){$V_L$}
\put(26,26){$V_{L\!-\!1}$}
\put(48.5,26){$V_2$}
\put(58.5,26){$V_1$}
\put(36,26){$\cdots\cdots$}
\end{picture}
}
\end{picture}

\noindent is expressed in terms of $R$ and 
$K$-matrices as follows \cite{Skl}: 
\begin{equation}
\begin{array}{rcl}
T_L (z_1 , z_2 )
&=&\mbox{Tr}_0\, 
K_+ (z_1 ) {\cal T}(z_1 , z_2 ) \\
{\cal T}(z_1 , z_2 ) &=&\mbox{Tr}_0\, 
{\cal T}(-z_1-z_2 )^{-1} K_- (z_1 ) 
{\cal T}(z_1 -z_2 ). 
\end{array}
\label{eq:T_L}
\end{equation}
Here 
$$
\begin{array}{rcl}
{\cal T}(z_1 -z_2 ) &=&
R_{01}^{V_{z_1},V_{z_2}} \cdots 
R_{0L}^{V_{z_1},V_{z_2}}\,\in
\mbox{End}\,(V_0 \otimes V_1 \otimes \cdots \otimes V_L ), \\
{\cal T}(-z_1 -z_2 )^{-1} &=&
R_{L0}^{V_{z_2},V_{-z_1}} \cdots 
R_{10}^{V_{z_2},V_{-z_1}}\,\in
\mbox{End}\,(V_0 \otimes V_1 \otimes \cdots \otimes V_L ), 
\end{array}
$$
are monodromy matrices satisfying 
\begin{equation}
R_{12}(z_1 -z_2 ){\cal T}_1 (z_1 ){\cal T}_2 (z_2 )
=
{\cal T}_2 (z_2 ){\cal T}_1 (z_1 )R_{12}(z_1 -z'_1 ), 
\label{eq:RTT} 
\end{equation}
and ${\mbox Tr}_0$ signifies the trance on the 
auxiliary space associated with the spectral 
parameters $z_1$ and $-z_1$. Note that the boundary 
monodromy matrix ${\cal T}(z,z')$ is a solution to 
the reflection equation: 
\begin{equation}
{\cal T}_2 (z'_1 , z_2 )R_{21}(z_1 +z_2 )
{\cal T}_1 (z_1 , z_2 )R_{12}(z_1 -z_2 )
=
R_{21}(z'_1 -z_1 ){\cal T}_1 (z_1 , z_2 )
R_{12}(z_1 +z'_1 ){\cal T}_2 (z'_1 , z_2 ). 
\label{eq:RTRT}
\end{equation}

\begin{prop} If one takes 
\begin{equation}
K_- (z)=K(z,v), ~~ K_+ (z) =K(-z-\tfrac{n}{2}w,v')\, 
\in \mbox{End}\,(V_0 ), 
\end{equation}
where $v$ and $v'$ are arbitrary parameters, 
the transfer matrices (\ref{eq:T_L}) commute 
each other \cite{Skl}: 
\begin{equation}
[T_L (z_1 , z_2 ), T_L (z'_1 , z_2 )]=0. 
\label{eq:com}
\end{equation}
\end{prop}
[Proof] From the crossing symmetry (\ref{eq:cross}) 
and the unitarity (\ref{eq:uni}) we have 
$$
\begin{array}{cl}
&T_L (z_1 , z_2 )T_L (z'_1 , z_2 ) \\
=& \mbox{Tr}_1 
K_1 (-z_1 -\tfrac{n}{2}w) {\cal T}_1 (z_1 , z_2 )\, 
\mbox{Tr}_2 
K_2 (-z'_1 -\tfrac{n}{2}w) {\cal T}_2 (z'_1 , z_2 ) \\
=& \mbox{Tr}_1 \mbox{Tr}_2 K_2 (-z'_1 -\tfrac{n}{2}w) 
K^{t_1}_1 (-z_1 -\tfrac{n}{2}w) 
{\cal T}^{t_1}_1 (z_1 , z_2 ){\cal T}_2 (z'_1 , z_2 ) \\
=& \mbox{Tr}_1 \mbox{Tr}_2 K_2 (-z'_1 -\tfrac{n}{2}w) 
K^{t_1}_1 (-z_1 -\tfrac{n}{2}w) 
R^{t_1}_{21} (-z_1 -z_2 -nw) R^{t_1}_{12} (z_1 +z_2 ) 
{\cal T}^{t_1}_1 (z_1 , z_2 ){\cal T}_2 (z'_1 , z_2 ) \\ 
=& \mbox{Tr}_1 \mbox{Tr}_2 K_2 (-z'_1 -\tfrac{n}{2}w) 
(R_{21} (-z_1 -z_2 -nw) K_1 (-z_1 -\tfrac{n}{2}w))^{t_1}
({\cal T}_1 (z_1 , z_2 )R_{12} (z_1 +z_2 ))^{t_1}
{\cal T}_2 (z'_1 , z_2 ) \\ 
=& \mbox{Tr}_1 \mbox{Tr}_2 K_2 (-z'_1 -\tfrac{n}{2}w) 
R_{21} (-z_1 -z_2 -nw) K_1 (-z_1 -\tfrac{n}{2}w)
R_{12} (z_2 -z_1 ) \\ 
\times & R_{21} (z_1 -z_2 )
{\cal T}_1 (z_1 , z_2 )R_{12} (z_1 +z_2 )
{\cal T}_2 (z'_1 , z_2 ), 
\end{array}
$$
where we use $\mbox{Tr}\,AB=\mbox{Tr}\,A^t B^t$. 
Furthermore, from (\ref{eq:RTRT}) we have 
$$
\begin{array}{cl}
=& \mbox{Tr}_1 \mbox{Tr}_2 R_{21} (z_2 -z_1 ) 
K_1 (-z_1 -\tfrac{n}{2}w) 
R_{12} (-z_1 -z_2 -nw) K_2 (-z'_1 -\tfrac{n}{2}w) 
\\ 
\times & {\cal T}_2 (z'_1 , z_2 ) R_{21} (z_1 +z_2 )
{\cal T}_1 (z_1 , z_2 )R_{12} (z_1 -z_2 )
\\
=& \mbox{Tr}_1 \mbox{Tr}_2 
K_1 (-z_1 -\tfrac{n}{2}w) 
(K_2 (-z'_1 -\tfrac{n}{2}w) R_{12} (-z_1 -z_2 -nw))^{t_2} 
(R_{21} (z_1 +z_2 ){\cal T}_2 (z'_1 , z_2 ))^{t_2} 
{\cal T}_1 (z_1 , z_2 )
\\
=& \mbox{Tr}_1 \mbox{Tr}_2 
K_1 (-z_1 -\tfrac{n}{2}w) 
(R_{12} (-z_1 -z_2 -nw)K_2 (-z'_1 -\tfrac{n}{2}w))^{t_2} 
({\cal T}_2 (z'_1 , z_2 )R_{21} (z_1 +z_2 ))^{t_2} 
{\cal T}_1 (z_1 , z_2 )
\\
=& \mbox{Tr}_1 \mbox{Tr}_2 
K_1 (-z_1 -\tfrac{n}{2}w) 
K^{t_2}_2 (-z'_1 -\tfrac{n}{2}w) 
{\cal T}^{t_2}_2 (z'_1 , z_2 ) 
{\cal T}_1 (z_1 , z_2 )
\\
=& T_L (z'_1 , z_2 )T_L (z_1 , z_2 ), 
\end{array}
$$
that implies the commutativity (\ref{eq:com}). $\Box$ 

\section{Boundary CTM bootstrap}

In this section we construct lattice realization 
of vertex operators and the boundary vacuum states 
for the boundary Belavin model. 

\subsection{Partition function}

Let us consider the inhomogeneous lattice 
${\cal L}_{LM}$ 
with $2M$ horizontal lines carrying alternating 
spectral parameters $z_1$ and $-z_1$ and 
$L(\equiv 0$ mod $n$) vertical lines 
carrying the spectral parameter $z_2$ 
as below: 

\begin{minipage}{8.5cm}
\unitlength 1mm
\begin{picture}(80,75)
\put(-2,10){\begin{picture}(101,0)
\put(10,20){\line(-1,-1){5}}
\put(5,15){\vector(1,-1){5}}
\put(10,10){\vector(1,0){60}}
\put(70,10){\line(1,1){5}}
\put(75,15){\vector(-1,1){5}}
\put(70,20){\vector(-1,0){60}}
\put(10,45){\line(-1,-1){5}}
\put(5,40){\vector(1,-1){5}}
\put(10,35){\vector(1,0){60}}
\put(70,35){\line(1,1){5}}
\put(75,40){\vector(-1,1){5}}
\put(70,45){\vector(-1,0){60}}
\put(20,0){\vector(0,1){55}}
\put(30,0){\vector(0,1){55}}
\put(50,0){\vector(0,1){55}}
\put(60,0){\vector(0,1){55}}
\put(5,0){\line(0,1){55}}
\put(5,15){\circle*{1}}
\put(5,40){\circle*{1}}
\put(75,0){\line(0,1){55}}
\put(75,15){\circle*{1}}
\put(75,40){\circle*{1}}
\put(39,7){$z_{1}$}
\put(37,17){$-z_{1}$}
\put(39,32){$z_{1}$}
\put(37,42){$-z_{1}$}
\put(54,26){$\vdots$}
\put(19,-4){$z_{2}$}
\put(29,-4){$z_{2}$}
\put(49,-4){$z_{2}$}
\put(59,-4){$z_{2}$}
\put(35,23){$\cdots\cdots$}
\put(61,4){$i$}
\put(67,7){$i$}
\put(67,21){$i$}
\put(67,32){$i$}
\put(67,46){$i$}
\put(61,49){$i$}
\put(51,4){$i_{-}$}
\put(54,7){$i_{-}$}
\put(61,14){$i_{-}$}
\put(54,21){$i_{-}$}
\put(54,32){$i_{-}$}
\put(61,39){$i_{-}$}
\put(54,46){$i_{-}$}
\put(51,49){$i_{-}$}
\put(51,14){$i_{--}$}
\put(51,39){$i_{--}$}
\put(11,7){$i$}
\put(18,14){$i$}
\put(11,21){$i$}
\put(11,32){$i$}
\put(18,39){$i$}
\put(11,46){$i$}
\put(16.5,4){$i_{+}$}
\put(24.5,7){$i_{+}$}
\put(31,14){$i_{+}$}
\put(24.5,21){$i_{+}$}
\put(24.5,32){$i_{+}$}
\put(31,39){$i_{+}$}
\put(24.5,46){$i_{+}$}
\put(16.5,49){$i_+$}
\put(31,4){$i_{++}$}
\put(31,49){$i_{++}$}
\put(20,5){\line(1,1){10}}
\put(30,15){\line(-1,1){7}}
\put(23,33){\line(1,1){7}}
\put(30,40){\line(-1,1){10}}
\put(22,23.85){$\vdots$}
\put(22,28.15){$\vdots$}
\end{picture}
}
\end{picture}
\end{minipage}
\begin{minipage}{6.5cm}
The lattice ${\cal L}_{LM}$ and the $i$-th 
ground state. \\ 
The arrows stand for the orientation of 
the spectral parameters. The dots $\bullet$'s 
stand for the boundary interaction $K(z)$. \\ 
For the sake of simplicity, 
we here denote the state $i\pm 1$ and 
$i\pm 2$ by $i_{\pm}$ and $i_{\pm\pm}$, 
respectively. \\
A zigzag line on which the state variables take 
$i+1$ is presented for transparency. 
\end{minipage}%

In this paper we restrict ourselves to the 
principal regime $0<t<q<r<u_\pm <1$, where 
$u_\pm =\exp \left( -\pi \sqrt{-1}(z_1 \pm z_2 )
\right)$. In this regime of parameters, the 
bulk Boltzmann weights of the type 
$R(z)^{j+1, j}_{j, j+1}$ 
dominates the others; and the boundary Boltzmann 
weight $K^i_i (z)$ is the largest among $K^j_i (z)$ 
for fixed $i$. 
Thus in the low temperature limit 
$t,q\rightarrow 0$, 
only the configuration such that the spin variables 
take the same value along the zigzag line 
(see the above figure) 
and increase by one in the direction from West to 
East, is possible. We call it a configuration of the 
ground state labeled by the boundary state $i\in \bz _n $.
Actually, the boundary weight $K^0_0 (z)$ 
(and $K^m_m (z)$ if $n$ is even) are the largest 
among $K^i_i (z)$. We therefore have only one 
real ground state for odd $n$ and two for even $n$. 
Nevertheless, we regard all $n$ kinds of configurations 
as the ground states. 

In what follows, we fix one of them (say, $i$) 
and define all the correlation functions 
in terms of the low-temperature series expansion 
(i.e., the formal power series of $t$ and $q$).   
Then the lowest order of them comes from 
the $i$-th ground state configuration. 
Furthermore, any finite order contribution 
is derived from 
the configurations which differ from 
that of the $i$-th ground state 
by altering a finite number of spins. 
It is equivalent to taking
the GNS representation obtained from the $i$-th 
ground state ($i$-th GNS representation) 
as the Hilbert space. 
It is expected that 
the correlation function defined in such a way 
is an analytic function which has a finite convergence radius 
if there exists the phase transition at a finite temperature. 

Following \cite{JKKMW} we conjecture the partition 
function $Z^{(i)}_{LM}(z_1 , z_2)$ of this model 
behaves in the thermodynamic limit $L, M \rightarrow 
\infty$ as 
\begin{equation}
\begin{array}{rcl}
\log Z^{(i)}_{LM}(z_1 , z_2) &\sim &
LM\left( \log\mu^{(i)}(z_1 -z_2 )+
\log\mu^{(i)}(z_1 +z_2 ) \right) \\
&+&
M \left( \log\nu^{(i)}(z_1 )+
\log\nu^{(i)}(-z_1 -\tfrac{n}{2}w) \right). 
\end{array}
\end{equation}
Here $\mu^{(i)}(z)$ is the partition function 
per cite for the bulk theory, and $\nu^{(i)}(z)$ 
is the that per boundary cite, which are normalized 
as follows: 
\begin{equation}
\mu^{(i)}(z)=1, ~~~
\nu^{(0)}(z)=1, ~~~
\nu^{(m)}(z)=1, \mbox{if $n$ is even}. 
\end{equation}

Next we consider the boundary CTM lattice as below: 

\begin{minipage}{8cm}
\unitlength 1mm
\begin{picture}(80,110)
\put(12,5){\begin{picture}(101,0)
\put(30,10){\vector(1,0){20}}
\put(50,10){\line(1,1){5}}
\put(55,15){\vector(-1,1){5}}
\put(50,20){\vector(-1,0){30}}
\put(10,30){\vector(1,0){40}}
\put(50,30){\line(1,1){5}}
\put(55,35){\vector(-1,1){5}}
\put(50,40){\vector(-1,0){50}}
\put(0,50){\vector(1,0){50}}
\put(50,50){\line(1,1){5}}
\put(55,55){\vector(-1,1){5}}
\put(50,60){\vector(-1,0){40}}
\put(20,70){\vector(1,0){30}}
\put(50,70){\line(1,1){5}}
\put(55,75){\vector(-1,1){5}}
\put(50,80){\vector(-1,0){20}}
\put(40,0){\vector(0,1){90}}
\put(30,10){\vector(0,1){70}}
\put(20,20){\vector(0,1){50}}
\put(10,30){\vector(0,1){30}}
\put(55,0){\line(0,1){90}}
\put(55,15){\circle*{1}}
\put(55,35){\circle*{1}}
\put(55,55){\circle*{1}}
\put(55,75){\circle*{1}}
\put(34,7){$z_{1}$}
\put(32,17){$-z_{1}$}
\put(34,27){$z_{1}$}
\put(32,37){$-z_{1}$}
\put(34,47){$z_{1}$}
\put(32,57){$-z_{1}$}
\put(34,67){$z_{1}$}
\put(32,77){$-z_{1}$}
\put(39,92){$z_{2}$}
\put(29,82){$z_{2}$}
\put(19,72){$z_{2}$}
\put(9,62){$z_{2}$}
\multiput(50,0)(0,2.0){45}{\line(0,1){1.0}}
\multiput(0,45)(2.0,0){28}{\line(1,0){1.0}}
\put(49,92){${}_i\langle B|$}
\put(50,-3.5){$|B\rangle_i$}
\put(5,12){$A^{(i)}_{SW}(z_1 , z_2 )$}
\put(5,78){$A^{(i)}_{NW}(z_1 , z_2 )$}
\end{picture}
}
\end{picture}
\end{minipage}
\begin{minipage}{6.5cm}
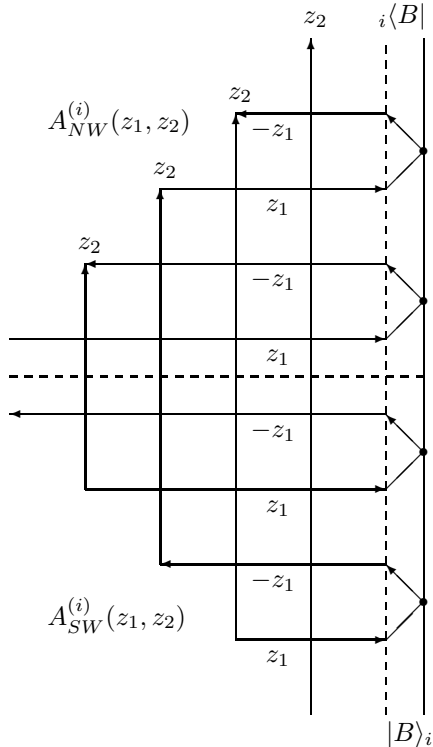
The inhomogeneous CTM lattice split 
into four sections. 
\end{minipage}%

We denote the SW and NW corner transfer matrices 
by $A^{(i)}_{SW}(z_1 , z_2 )$ and 
$A^{(i)}_{NW}(z_1 , z_2 )$, respectively; and 
also denote the upper and lower lines of 
$K(z)$ by ${}_i\langle B|$ and $|B\rangle_i$, 
respectively. Let 
\begin{equation}
\begin{array}{rcl}
{\cal H}^{(i)}&:=&\{ 
\cdots \otimes v_{p(3)}\otimes v_{p(2)}\otimes 
v_{p(1)} | p(j)\in \bz_n , p(j)=i+1-j \mbox{
(mod $n$) for $j\gg 1$} \}, \\
{\cal \bar{H}}^{(i)}&:=&\{ 
\cdots \otimes v_{p(3)}\otimes v_{p(2)}\otimes 
v_{p(1)} | p(j)\in \bz_n , p(j)=i \mbox{
(mod $n$) for $j\gg 1$} \}, 
\end{array}
\end{equation}
and ${\cal H}^{*(i)}$ and ${\cal \bar{H}}^{*(i)}$ 
be their dual spaces. Then in the infinite lattice 
limit we conclude that $|B\rangle_i \in {\cal \bar{H}}$, 
${}_i\langle B| \in {\cal \bar{H}}^{*(i)}$, and 
\begin{equation}
\begin{array}{rl}
A^{(i)}_{SW}(z_1 , z_2 ): & {\cal \bar{H}}^{(i)} 
\longrightarrow {\cal H}^{(i)}, \\
A^{(i)}_{NW}(z_1 , z_2 ): & {\cal H}^{(i)} 
\longrightarrow {\cal \bar{H}}^{*(i)}. 
\end{array}
\end{equation}
The partition function is given as follows: 
\begin{equation}
Z^{(i)}(z_1 , z_2 )={}_i\langle B|
A^{(i)}_{NW}(z_1 , z_2 )A^{(i)}_{SW}(z_1 , z_2 )
|B\rangle_i . 
\label{eq:Z}
\end{equation}

\subsection{Vertex operators}

Let us introduce the type I vertex operators 

\unitlength 1mm
\begin{picture}(140,90)(0,-20)
\put(15,60){\vector(1,0){45}}
\put(30,55){\vector(0,1){10}}
\put(40,55){\vector(0,1){10}}
\put(50,55){\vector(0,1){10}}
\put(11.5,59.5){$z_1$}
\put(56,61.5){$j$}
\put(29,52){$z_2$}
\put(39,52){$z_2$}
\put(49,52){$z_2$}
\put(22,61){$\cdots$}
\put(65,59){$=\phi^j_{(i-1,i)}(z_1 -z_2 ): 
{\cal H}^{(i)}\longrightarrow 
{\cal H}^{(i-1)},$}
\put(60,40){\vector(-1,0){45}}
\put(30,35){\vector(0,1){10}}
\put(40,35){\vector(0,1){10}}
\put(50,35){\vector(0,1){10}}
\put(11.5,39.5){$z_1$}
\put(56,41.5){$j$}
\put(29,32){$z_2$}
\put(39,32){$z_2$}
\put(49,32){$z_2$}
\put(22,41){$\cdots$}
\put(65,39){$=\phi_j^{(i+1,i)}(z_2 -z_1 ): 
{\cal H}^{(i)}\longrightarrow 
{\cal H}^{(i+1)},$}
\put(15,20){\vector(1,0){45}}
\put(30,15){\vector(0,1){10}}
\put(40,15){\vector(0,1){10}}
\put(50,15){\vector(0,1){10}}
\put(11.5,19.5){$z_1$}
\put(56,21.5){$j^*$}
\put(29,12){$z_2$}
\put(39,12){$z_2$}
\put(49,12){$z_2$}
\put(22,21){$\cdots$}
\put(65,19){$=\phi^{*j}_{(i+1,i)}(z_1 -z_2 ): 
{\cal H}^{(i)}\longrightarrow 
{\cal H}^{(i+1)},$}
\put(60,0){\vector(-1,0){45}}
\put(30,-5){\vector(0,1){10}}
\put(40,-5){\vector(0,1){10}}
\put(50,-5){\vector(0,1){10}}
\put(11.5,-0.5){$z_1$}
\put(56,1.5){$j^*$}
\put(29,-8){$z_2$}
\put(39,-8){$z_2$}
\put(49,-8){$z_2$}
\put(22,1){$\cdots$}
\put(65,-1){$=\phi^{*(i-1,i)}_j (z_2 -z_1 ): 
{\cal H}^{(i)}\longrightarrow 
{\cal H}^{(i-1)},$}
\end{picture}

\noindent where the sub/superscripts $(i\pm 1,i)$ 
specify the spaces intertwined by the vertex operators. 
We often suppress these sub/superscripts when we 
have no fear of confusion. 

It follows from 
the Yang--Baxter equation that these vertex operators 
satisfy the following commutation relations 
\cite{JM,JKKMW}: 
\begin{equation}
\begin{array}{rcl}
\phi^{j_2}(z_2 )\phi^{j_1}(z_1 )&=&
\displaystyle\sum_{j'_1,j'_2} 
(R^{V_{z_1},V_{z_2}})^{j_1 j_2}_{j'_1 j'_2}
\phi^{j'_1}(z_1 )\phi^{j'_2}(z_2 ), \\
\phi^{*j_2}(z_2 )\phi^{j_1}(z_1 )&=&
\displaystyle\sum_{j'_1,j'_2} 
(R^{V_{z_1},V^*_{z_2}})^{j_1 j_2}_{j'_1 j'_2}
\phi^{j'_1}(z_1 )\phi^{*j'_2}(z_2 ), \\
\phi^{*j_2}(z_2 )\phi^{*j_1}(z_1 )&=&
\displaystyle\sum_{j'_1,j'_2} 
(R^{V^*_{z_1},V^*_{z_2}})^{j_1 j_2}_{j'_1 j'_2}
\phi^{*j'_1}(z_1 )\phi^{*j'_2}(z_2 ). 
\end{array}
\label{eq:CR-VO}
\end{equation}
Furthermore, the unitarity relations for 
$R$-matrices imply the inversion relation of 
the vertex operators: 
\begin{equation}
\sum_{j} \phi_j (-z)\phi^j (z)=1, ~~~~
\sum_{j} \phi^*_j (-z)\phi^{*j}(z)=1. 
\label{eq:inv-phi}
\end{equation}
{}From the crossing symmetry we have 
\begin{equation}
\phi^{*j}(z)=\phi_j (-z-\tfrac{n}{2}w), ~~~~
\phi^*_{j}(-z)=\phi^j (z-\tfrac{n}{2}w). 
\label{eq:x-VO}
\end{equation}
Using these vertex operators, the transfer matrix 
for the semi-infinite 
lattice is defined as follow: 
\begin{equation}
\begin{array}{rcl}
T_B (z_1 , z_2 ) 
&=&\displaystyle\sum_{j,k} \phi_j (z_1 +z_2 ) 
K^j_k (z_1 ) \phi^k (z_1 -z_2 ) \\
&=&\displaystyle\sum_{j,k} \phi^{*j} (-z_1 -
\tfrac{n}{2}w-z_2 ) K^j_k (z_1 ) \phi^k (z_1 -z_2 ). 
\end{array}
\end{equation}
If the $i$-th vauuam states $|\mbox{vac}\rangle _i$ 
and ${}_i\langle \mbox{vac}|$ satisfy the following 
reflection properties: 
\begin{equation}
\begin{array}{rcl}
\displaystyle\sum_{k} K^j_k (z) 
\phi^k (z) |\mbox{vac}\rangle _i 
&=&\nu^{(i)}(z) 
\phi^j (-z)|\mbox{vac}\rangle _i , \\ 
{}_i\langle \mbox{vac}| \displaystyle\sum_{k} 
\phi_k (z) K_j^k (z) 
&=&\nu^{(i)}(z) 
{}_i\langle \mbox{vac}| \phi_j (-z), 
\label{eq:Ref-Pro}
\end{array}
\end{equation}
these vacuums are the eigenstates of $T_B (z,0)$ 
associated with the 
eigenvalues $\nu^{(i)}(z)$, respectively: 
$$
T_B (z, 0) |\mbox{vac}\rangle _i =
\nu^{(i)}(z) |\mbox{vac}\rangle _i , ~~~~
{}_i\langle \mbox{vac}|T_B (z, 0) =\nu^{(i)}(z) 
{}_i\langle \mbox{vac}|. 
$$

For $n=2$, it is suffices to consider only 
two types vertex operators $\phi^j (z)$ and 
$\phi_j (z)$ because of 
$\phi^{*j}(z)=\phi_{1-j} (-z-w)$ and 
$\phi^{*}_j (z)=\phi^{1-j} (-z-w)$ \cite{JKKMW}. 
Furthermore, from $T_B (z_1 , z_2 ) 
=T_B (-z_1 -w, z_2 )$ for $n=2$, we have 
\begin{equation}
\begin{array}{cl}
&\displaystyle\sum_{j,k} \phi^{1-j} 
(-z_1 -w-z_2 ) 
K^j_k (z_1 ) \phi^k (z_1 -z_2 ) \\
=&\displaystyle\sum_{j',k'} \phi^{1-k'} (z_1 -z_2 ) 
K^{k'}_{j'} (-z_1 -w) \phi^{j'} (-z_1 -w-z_2 ) \\
=&\displaystyle\sum_{j,k\atop j',k'} 
R(-2z_1 -w)^{j'\,1-k'}_{1-j\,k}
\phi^{1-j} (-z_1 -w-z_2 ) \phi^k (z_1 -z_2 ) 
K^{j'}_{k'} (-z_1 -w), 
\end{array}
\label{eq:der}
\end{equation}
which implies the boundary crossing symmetry (\ref{eq:b-x}). 

The crucial point in (\ref{eq:der}) 
consists in the self-duality $\phi^*_j (z)=\phi^{1-j}(z)$ 
for $n=2$. Thus the boundary crossing symmetry (\ref{eq:b-x}) 
does not have a simple generalization for $n>2$. We should 
rather regard the RHS of (\ref{eq:b-x}) for general $n$ as 
the definition of the dual $K$-matrix. In order to see that, 
let us repeat the reduction (\ref{eq:der}) for general $n$. 
Using eqs. (\ref{eq:x-VO}), (\ref{eq:Ref-Pro}), 
(\ref{eq:CR-VO}) and (\ref{eq:inv-phi}) we have 
$$
\begin{array}{rcl}
\nu ^{(i)}(z)&=&\displaystyle\sum_{j',k'} 
{}_i\langle \mbox{vac}| \phi^{*k'}(-z-\tfrac{n}{2}w) 
K_{j'}^{k'}(z) \phi^{j'}(z)|\mbox{vac}\rangle_i \\
&=&\displaystyle\sum_{j,k\atop j',k'} 
{}_i\langle \mbox{vac}| \phi^{j}(z) 
(R^{V_z , V^*_{-z-nw/2}})^{j'k'}_{jk}
K_{j'}^{k'}(z) \phi^{*k}(-z-\tfrac{n}{2}w) 
|\mbox{vac}\rangle_i \\
&=&\displaystyle\sum_{j,k\atop j',k'} 
{}_i\langle \mbox{vac}| \phi^*_{j}(-z-\tfrac{n}{2}w) 
(R^{V_z , V^*_{-z-nw/2}})^{j'k'}_{jk}
K_{j'}^{k'}(z) \phi^{*k}(-z-\tfrac{n}{2}w) 
|\mbox{vac}\rangle_i . 
\end{array}
$$
Thus, if we define the dual $K$-matrix by 
\begin{equation}
K^* (-z-\tfrac{n}{2}w)^j_k :=
\sum_{j',k'} 
(R^{V_z , V^*_{-z-nw/2}})^{j'k'}_{jk}
K(z)_{j'}^{k'}, 
\label{df:K^*}
\end{equation}
then the following dual reflection properties hold: 
\begin{equation}
\begin{array}{rcl}
\displaystyle\sum_{k} K^* (z)^j_k 
\phi^{*k}(z) |\mbox{vac}\rangle _i 
&=&\nu^{(i)}(-z-\tfrac{n}{2}w) 
\phi^{*j}(-z)|\mbox{vac}\rangle _i , \\ 
{}_i\langle \mbox{vac}| \displaystyle\sum_{k} 
\phi^*_k (z) K^* (z)_j^k 
&=&\nu^{(i)}(-z-\tfrac{n}{2}w) 
{}_i\langle \mbox{vac}| \phi^*_j (-z).
\end{array} 
\label{eq:Ref-Pro*}
\end{equation}

The associativity condition of the algebra 
(\ref{eq:CR-VO}) and (\ref{eq:Ref-Pro*}) implies 
the reflection equations involving $K^*$-matrices: 
\begin{equation}
\begin{array}{rcl}
K_2 (z_2 )R_{21}^{V_{z_2},V^*_{-z_1}}
K^*_1 (z_1 )R_{12}^{V^*_{z_1},V_{z_2}}
&=&
R_{21}^{V_{-z_2},V^*_{-z_1}}K^*_1 (z_1 )
R_{12}^{V^*_{z_1},V_{-z_2}}
K_2 (z_2 ), \\
K^*_2 (z_2 )R_{21}^{V^*_{z_2},V^*_{-z_1}}
K^*_1 (z_1 )R_{12}^{V^*_{z_1},V^*_{z_2}}
&=&
R_{21}^{V^*_{-z_2},V^*_{-z_1}}K^*_1 (z_1 )
R_{12}^{V^*_{z_1},V^*_{-z_2}}
K^*_2 (z_2 ). 
\end{array}
\end{equation}

\subsection{Derivation of the reflection properties}

In this subsection we derive the reflection properties 
(\ref{eq:Ref-Pro},\ref{eq:Ref-Pro*}). For that purpose 
we introduce following further 
two types of vertex operators: 

~

\unitlength 1mm
\begin{picture}(140,55)(0,-20)
\put(-10,-55){\begin{picture}(101,0)
\put(15,45){\vector(0,1){45}}
\put(20,80){\vector(-1,0){10}}
\put(10,73){\vector(1,0){10}}
\put(20,66){\vector(-1,0){10}}
\put(10,59){\vector(1,0){10}}
\put(13,50){$\vdots$}
\put(16.5,85){$j$}
\put(4,79.5){$-z_1$}
\put(6.5,72.5){$z_1$}
\put(4,65.5){$-z_1$}
\put(6.5,58.5){$z_1$}
\put(13.8,42){$z_3$}
\put(25,68){$=\varphi^j_{(i-1,i)} (z_1 , z_3 ): 
{\cal \bar{H}}^{(i)}\longrightarrow 
{\cal \bar{H}}^{(i-1)},$}
\end{picture}
}
\put(70,-55){\begin{picture}(101,0)
\put(15,45){\vector(0,1){45}}
\put(20,76){\vector(-1,0){10}}
\put(10,69){\vector(1,0){10}}
\put(20,62){\vector(-1,0){10}}
\put(10,55){\vector(1,0){10}}
\put(13,82){$\vdots$}
\put(16.5,48.5){$j$}
\put(4,75.5){$-z_1$}
\put(6.5,68.5){$z_1$}
\put(4,61.5){$-z_1$}
\put(6.5,54.5){$z_1$}
\put(13.8,42){$z_3$}
\put(25,68){$=\varphi_{j}^{(i-1,i)} (z_1 , z_3 ): 
{\cal \bar{H}}^{(*i)}\longrightarrow 
{\cal \bar{H}}^{(*i-1)}.$}
\end{picture}
}
\end{picture}

\noindent where the sub/superscripts $(i\pm 1,i)$ 
specify 
the spaces intertwined by the vertex operators. 
Hereafter we also suppress these sub/superscripts. 

{}From the reflection equation (\ref{eq:RE})

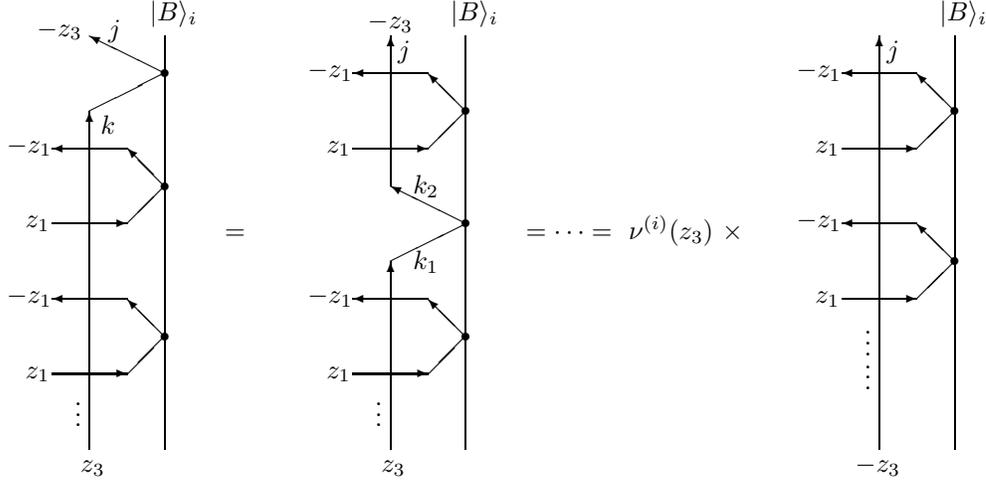
\unitlength 1mm
\begin{picture}(140,70)(0,35)
\put(15,40){\vector(0,1){45}}
\put(20,80){\vector(-1,0){10}}
\put(10,70){\vector(1,0){10}}
\put(20,60){\vector(-1,0){10}}
\put(10,50){\vector(1,0){10}}
\put(13,43){$\vdots$}
\put(16.5,82){$k$}
\put(4,79.5){$-z_1$}
\put(6.5,69.5){$z_1$}
\put(4,59.5){$-z_1$}
\put(6.5,49.5){$z_1$}
\put(13.8,37){$z_3$}
\put(25,75){\vector(-1,1){5}}
\put(20,70){\line(1,1){5}}
\put(25,55){\vector(-1,1){5}}
\put(20,50){\line(1,1){5}}
\put(25,40){\line(0,1){55}}
\put(25,75){{\circle*{1}}}
\put(25,55){{\circle*{1}}}
\put(23,97){$|B\rangle_i$}
\put(25,90){\vector(-2,1){10}}
\put(15,85){\line(2,1){10}}
\put(17.5,94.8){$j$}
\put(8,95){$-z_3$}
\put(25,90){{\circle*{1}}}
\put(33,68){$=$}
\put(55,40){\vector(0,1){25}}
\put(55,75){\vector(0,1){20}}
\put(60,90){\vector(-1,0){10}}
\put(50,80){\vector(1,0){10}}
\put(60,60){\vector(-1,0){10}}
\put(50,50){\vector(1,0){10}}
\put(53,43){$\vdots$}
\put(58,74){$k_2$}
\put(58,64){$k_1$}
\put(44,89.5){$-z_1$}
\put(46.5,79.5){$z_1$}
\put(44,59.5){$-z_1$}
\put(46.5,49.5){$z_1$}
\put(53.8,37){$z_3$}
\put(65,85){\vector(-1,1){5}}
\put(60,80){\line(1,1){5}}
\put(65,55){\vector(-1,1){5}}
\put(60,50){\line(1,1){5}}
\put(65,40){\line(0,1){55}}
\put(65,85){{\circle*{1}}}
\put(65,55){{\circle*{1}}}
\put(63,97){$|B\rangle_i$}
\put(65,70){\vector(-2,1){10}}
\put(55,65){\line(2,1){10}}
\put(56,92){$j$}
\put(52,96){$-z_3$}
\put(65,70){{\circle*{1}}}
\put(73,68){$= \cdots =\,\,
\nu^{(i)}(z_3 )\,\,\times$}
\put(120,40){\vector(0,1){55}}
\put(125,90){\vector(-1,0){10}}
\put(115,80){\vector(1,0){10}}
\put(125,70){\vector(-1,0){10}}
\put(115,60){\vector(1,0){10}}
\put(118,52.5){$\vdots$}
\put(118,48){$\vdots$}
\put(109,89.5){$-z_1$}
\put(111.5,79.5){$z_1$}
\put(109,69.5){$-z_1$}
\put(111.5,59.5){$z_1$}
\put(116.8,37){$-z_3$}
\put(130,85){\vector(-1,1){5}}
\put(125,80){\line(1,1){5}}
\put(130,65){\vector(-1,1){5}}
\put(125,60){\line(1,1){5}}
\put(130,40){\line(0,1){55}}
\put(130,85){{\circle*{1}}}
\put(130,65){{\circle*{1}}}
\put(128,97){$|B\rangle_i$}
\put(121,92){$j$}
\end{picture}

\noindent we have the following relation: 
\begin{eqnarray}
\sum_{k} K(z_3 )^j_{k}\varphi^{k}
(z_1 , z_3) |B\rangle_i &=&
\nu^{(i)}(z_3 )\varphi^j (z_1 , -z_3 )
|B\rangle_i . \label{eq:KP^|B>} 
\end{eqnarray}
By similar argument we have 
\begin{eqnarray}
\sum_{k} {}_i \langle B| \varphi_{k}
(z_1 , -z_3) K(z_3 )_j^{k} &=&
\nu^{(i)}(z_3 ){}_i \langle B|\varphi_{j}(z_1 , z_3 ), 
\label{eq:<B|P_K} 
\end{eqnarray}

Furthermore, we have the relations
\begin{eqnarray}
A^{(i-1)}_{SW} (z_1 , z_2 ) \varphi^j 
(z_1 , z_3 )|B\rangle_i &=& 
\phi^j (z_3 -z_2 )A^{(i)}_{SW} (z_1 , z_2 )
|B\rangle_i , \label{eq:SWP^} 
\\
{}_i\langle B| \varphi_{j} (z_1 , z_3 ) 
A^{(i-1)}_{NW} (z_1 , z_2 ) &=& {}_i\langle B| 
A^{(i)}_{NW} (z_1 , z_2 )\phi_{j} (z_2 -z_3 ). 
\label{eq:NWP^} 
\end{eqnarray}
These are based on the unitarity and Yang-Baxter 
relation of $R$-matrix in the thermodynamic limit. 
The unitarity (\ref{eq:cross}) allows us to obtain 

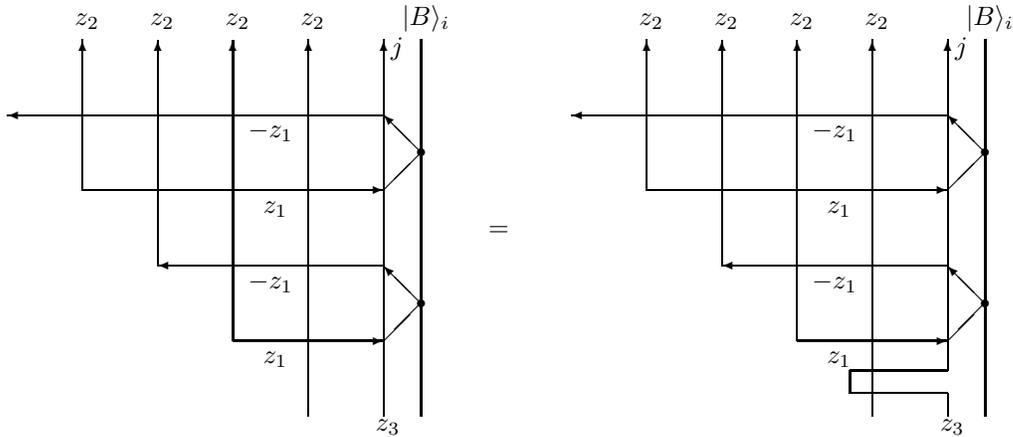
\unitlength 1mm
\begin{picture}(140,65)(-5,-5)
\put(30,10){\vector(1,0){20}}
\put(50,10){\line(1,1){5}}
\put(55,15){\vector(-1,1){5}}
\put(50,20){\vector(-1,0){30}}
\put(10,30){\vector(1,0){40}}
\put(50,30){\line(1,1){5}}
\put(55,35){\vector(-1,1){5}}
\put(50,40){\vector(-1,0){50}}
\put(40,0){\vector(0,1){50}}
\put(30,10){\vector(0,1){40}}
\put(20,20){\vector(0,1){30}}
\put(10,30){\vector(0,1){20}}
\put(55,0){\line(0,1){50}}
\put(55,15){\circle*{1}}
\put(55,35){\circle*{1}}
\put(34,7){$z_{1}$}
\put(32,17){$-z_{1}$}
\put(34,27){$z_{1}$}
\put(32,37){$-z_{1}$}
\put(39,52){$z_{2}$}
\put(29,52){$z_{2}$}
\put(19,52){$z_{2}$}
\put(9,52){$z_{2}$}
\put(52.5,52){$|B\rangle_i$}
\put(50,0){\vector(0,1){50}}
\put(49,-2){$z_{3}$}
\put(51,48){$j$}
\put(64,24){$=$}
\put(105,10){\vector(1,0){20}}
\put(125,10){\line(1,1){5}}
\put(130,15){\vector(-1,1){5}}
\put(125,20){\vector(-1,0){30}}
\put(85,30){\vector(1,0){40}}
\put(125,30){\line(1,1){5}}
\put(130,35){\vector(-1,1){5}}
\put(125,40){\vector(-1,0){50}}
\put(115,0){\vector(0,1){50}}
\put(105,10){\vector(0,1){40}}
\put(95,20){\vector(0,1){30}}
\put(85,30){\vector(0,1){20}}
\put(130,0){\line(0,1){50}}
\put(130,15){\circle*{1}}
\put(130,35){\circle*{1}}
\put(109,7){$z_{1}$}
\put(107,17){$-z_{1}$}
\put(109,27){$z_{1}$}
\put(107,37){$-z_{1}$}
\put(114,52){$z_{2}$}
\put(104,52){$z_{2}$}
\put(94,52){$z_{2}$}
\put(84,52){$z_{2}$}
\put(127.5,52){$|B\rangle_i$}
\put(125,0){\line(0,1){3}}
\put(125,3){\line(-1,0){13}}
\put(112,3){\line(0,1){3}}
\put(112,6){\line(1,0){13}}
\put(125,6){\vector(0,1){44}}
\put(124,-2){$z_{3}$}
\put(126,48){$j$}
\end{picture}

Using the Yang--Baxter equation (\ref{YBE}) we get 

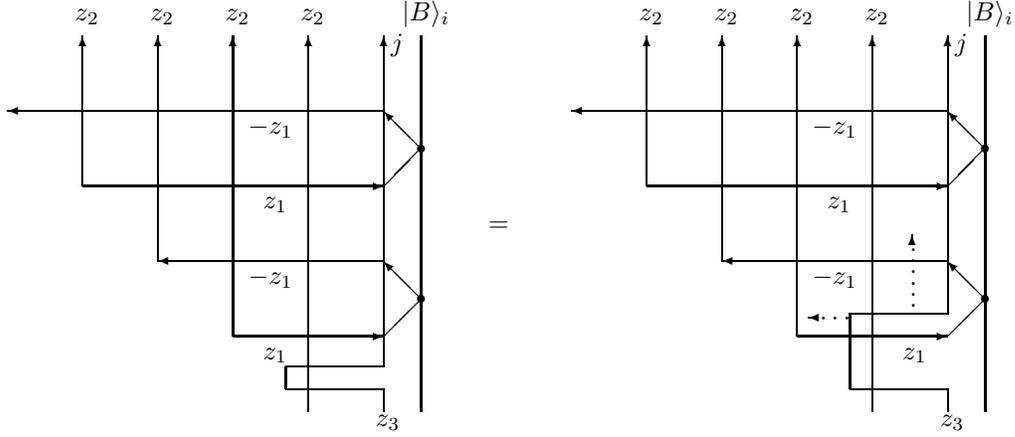
\unitlength 1mm
\begin{picture}(140,65)(-5,-5)
\put(30,10){\vector(1,0){20}}
\put(50,10){\line(1,1){5}}
\put(55,15){\vector(-1,1){5}}
\put(50,20){\vector(-1,0){30}}
\put(10,30){\vector(1,0){40}}
\put(50,30){\line(1,1){5}}
\put(55,35){\vector(-1,1){5}}
\put(50,40){\vector(-1,0){50}}
\put(40,0){\vector(0,1){50}}
\put(30,10){\vector(0,1){40}}
\put(20,20){\vector(0,1){30}}
\put(10,30){\vector(0,1){20}}
\put(55,0){\line(0,1){50}}
\put(55,15){\circle*{1}}
\put(55,35){\circle*{1}}
\put(34,7){$z_{1}$}
\put(32,17){$-z_{1}$}
\put(34,27){$z_{1}$}
\put(32,37){$-z_{1}$}
\put(39,52){$z_{2}$}
\put(29,52){$z_{2}$}
\put(19,52){$z_{2}$}
\put(9,52){$z_{2}$}
\put(52.5,52){$|B\rangle_i$}
\put(50,0){\line(0,1){3}}
\put(50,3){\line(-1,0){13}}
\put(37,3){\line(0,1){3}}
\put(37,6){\line(1,0){13}}
\put(50,6){\vector(0,1){44}}
\put(49,-2){$z_{3}$}
\put(51,48){$j$}
\put(64,24){$=$}
\put(105,10){\vector(1,0){20}}
\put(125,10){\line(1,1){5}}
\put(130,15){\vector(-1,1){5}}
\put(125,20){\vector(-1,0){30}}
\put(85,30){\vector(1,0){40}}
\put(125,30){\line(1,1){5}}
\put(130,35){\vector(-1,1){5}}
\put(125,40){\vector(-1,0){50}}
\put(115,0){\vector(0,1){50}}
\put(105,10){\vector(0,1){40}}
\put(95,20){\vector(0,1){30}}
\put(85,30){\vector(0,1){20}}
\put(130,0){\line(0,1){50}}
\put(130,15){\circle*{1}}
\put(130,35){\circle*{1}}
\put(119,7){$z_{1}$}
\put(107,17){$-z_{1}$}
\put(109,27){$z_{1}$}
\put(107,37){$-z_{1}$}
\put(114,52){$z_{2}$}
\put(104,52){$z_{2}$}
\put(94,52){$z_{2}$}
\put(84,52){$z_{2}$}
\put(127.5,52){$|B\rangle_i$}
\put(125,0){\line(0,1){3}}
\put(125,3){\line(-1,0){13}}
\put(112,3){\line(0,1){10}}
\put(112,13){\line(1,0){13}}
\put(125,13){\vector(0,1){37}}
\put(124,-2){$z_{3}$}
\put(126,48){$j$}
\put(120,14){$\vdots$}
\put(120,18.5){$\vdots$}
\put(120.3,23){\vector(0,1){0.5}}
\put(108,11.5){$\cdots$}
\put(107,12.5){\vector(-1,0){0.5}}
\end{picture}

\noindent By successive use of the YBE and the unitarity 
we can bring the line associated with the spectral parameter 
$z_3$ to the directions pointed by dotted lines in the above 
figure as far as we like. Thus we find 

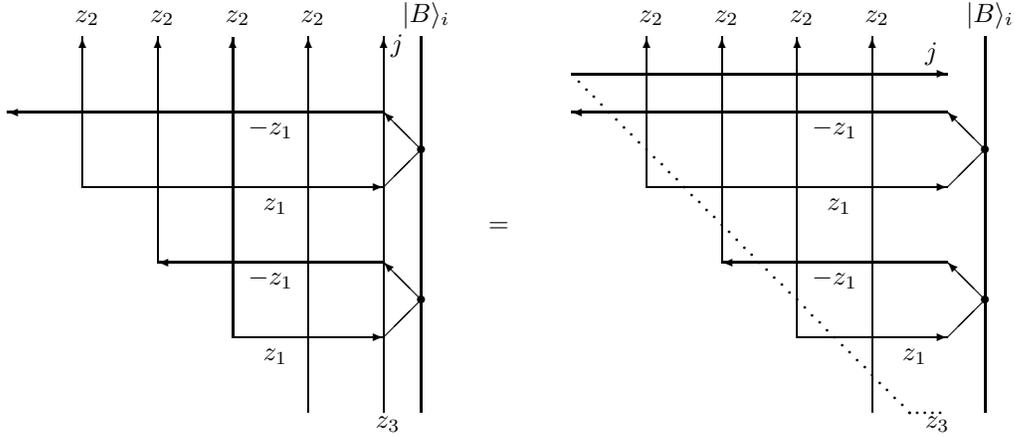
\unitlength 1mm
\begin{picture}(140,65)(-5,-5)
\put(30,10){\vector(1,0){20}}
\put(50,10){\line(1,1){5}}
\put(55,15){\vector(-1,1){5}}
\put(50,20){\vector(-1,0){30}}
\put(10,30){\vector(1,0){40}}
\put(50,30){\line(1,1){5}}
\put(55,35){\vector(-1,1){5}}
\put(50,40){\vector(-1,0){50}}
\put(40,0){\vector(0,1){50}}
\put(30,10){\vector(0,1){40}}
\put(20,20){\vector(0,1){30}}
\put(10,30){\vector(0,1){20}}
\put(55,0){\line(0,1){50}}
\put(55,15){\circle*{1}}
\put(55,35){\circle*{1}}
\put(34,7){$z_{1}$}
\put(32,17){$-z_{1}$}
\put(34,27){$z_{1}$}
\put(32,37){$-z_{1}$}
\put(39,52){$z_{2}$}
\put(29,52){$z_{2}$}
\put(19,52){$z_{2}$}
\put(9,52){$z_{2}$}
\put(52.5,52){$|B\rangle_i$}
\put(50,0){\vector(0,1){50}}
\put(49,-2){$z_{3}$}
\put(51,48){$j$}
\put(64,24){$=$}
\put(105,10){\vector(1,0){20}}
\put(125,10){\line(1,1){5}}
\put(130,15){\vector(-1,1){5}}
\put(125,20){\vector(-1,0){30}}
\put(85,30){\vector(1,0){40}}
\put(125,30){\line(1,1){5}}
\put(130,35){\vector(-1,1){5}}
\put(125,40){\vector(-1,0){50}}
\put(115,0){\vector(0,1){50}}
\put(105,10){\vector(0,1){40}}
\put(95,20){\vector(0,1){30}}
\put(85,30){\vector(0,1){20}}
\put(130,0){\line(0,1){50}}
\put(130,15){\circle*{1}}
\put(130,35){\circle*{1}}
\put(119,7){$z_{1}$}
\put(107,17){$-z_{1}$}
\put(109,27){$z_{1}$}
\put(107,37){$-z_{1}$}
\put(114,52){$z_{2}$}
\put(104,52){$z_{2}$}
\put(94,52){$z_{2}$}
\put(84,52){$z_{2}$}
\put(127.5,52){$|B\rangle_i$}
\multiput(120,0)(1.0,0){5}{\circle*{0.5}}
\multiput(120,0)(-1.0,1.0){45}{\circle*{0.5}}
\put(75,45){\vector(1,0){50}}
\put(122,-2){$z_{3}$}
\put(122,47){$j$}
\end{picture}

\noindent Those manipulation implies (\ref{eq:SWP^}) 
because the contribution of Boltzmann weights along 
the tail graphically represented in the figure by 
the dotted line is unity in the thermodynamic limit. 
The relation (\ref{eq:NWP^}) can be similarly obtained. 

Applying $A^{(i-1)}_{SW}(z_1 , z_2 )$ 
(resp. $A^{(i-1)}_{NW}(z_1 , z_2 )$) from the 
left (resp. right) to both sides of 
(\ref{eq:KP^|B>}) (resp. (\ref{eq:<B|P_K})) and using 
(\ref{eq:SWP^}) (resp. (\ref{eq:NWP^})) we obtain 
\begin{eqnarray}
\sum_{k} K(z_3 )^j_{k}\phi^{k}
(z_3 -z_2 ) A^{(i)}_{SW} (z_1 , z_2 )|B\rangle_i 
&=& \nu^{(i)}(z_3 )\phi^j (-z_3 -z_2 )
A^{(i)}_{SW} (z_1 , z_2 )|B\rangle_i , 
\label{eq:RE-P}
\\
\sum_{k} {}_i\langle B| A^{(i)}_{NW} (z_1 , z_2 )
\phi_{k}(z_2 +z_3 ) K(z_3 )_j^{k}&=&
\nu^{(i)}(z_3 ){}_i\langle B|A^{(i)}_{NW} (z_1 , z_2 )
\phi_{j} (z_2 -z_3 ). 
\label{eq:RE-P*} 
\end{eqnarray}
Taking account of (\ref{eq:RE-P}) and (\ref{eq:RE-P*}) 
with (\ref{eq:Ref-Pro}) we find the following identification 
\begin{equation}
|\mbox{vac}\rangle_i 
=A^{(i)}_{SW} (z_1 , z_2 =0)|B\rangle_i , 
~~~~
{}_i\langle \mbox{vac} |
={}_i\langle B|A^{(i)}_{SW} (z_1 , z_2 =0). 
\label{df:vac}
\end{equation}
{}From the identification (\ref{df:vac}) and 
the definition of the dual $K$-matrix (\ref{df:K^*}) 
we obtain 
\begin{eqnarray}
\sum_{k} K^* (z_3 )^j_{k}\phi^{*k}
(z_3 -z_2 ) A^{(i)}_{SW} (z_1 , z_2 )|B\rangle_i 
&=&
\nu^{(i)}(-z_3 -\tfrac{n}{2}w)\phi^{*j} (-z_3 -z_2 )
A^{(i)}_{SW} (z_1 , z_2 )|B\rangle_i , 
\label{eq:RE-P_} \\
\sum_{k} {}_i\langle B| A^{(i)}_{NW} (z_1 , z_2 )
\phi^{*}_k(z_2 +z_3 ) K^*(z_3 )_j^{k}&=&
\nu^{(i)}(-z_3 -\tfrac{n}{2}w)
{}_i\langle B|A^{(i)}_{NW} (z_1 , z_2 )
\phi^{*}_j (z_2 -z_3 ). \label{eq:RE-P_*}
\end{eqnarray}

\section{Correlation functions and difference equations}

The relations appeared in the previous section are 
not rigorous because all the objects are defined on 
the infinite lattice. Nevertheless we assume that eqs. 
(4.1--\ref{eq:RE-P_*}) are exactly correct on the basis 
of the CTM bootstrap method, which is supported by 
some numerical calculations \cite{ESM} and consistency 
with the vertex operator method \cite{JM}. 

\subsection{Local state probabilities}

Let us consider the correlation function 
on the dislocated CTM lattice 

\unitlength 1mm
\begin{picture}(140,120)
\put(10,110){$G^{(i)}_N (z, z' | 
z'_1 , \cdots , z'_N , z_N , \cdots , 
z_1 )^{j'_1 , \cdots j'_N , j_N , \cdots , j_1}$}
\put(8,50){$=$}
\put(15,5){
\begin{picture}(101,0)
\put(30,10){\vector(1,0){40}}
\put(70,10){\line(1,1){5}}
\put(75,15){\vector(-1,1){5}}
\put(70,20){\vector(-1,0){50}}
\put(10,30){\vector(1,0){60}}
\put(70,30){\line(1,1){5}}
\put(75,35){\vector(-1,1){5}}
\put(70,40){\vector(-1,0){70}}
\put(0,55){\vector(1,0){70}}
\put(70,55){\line(1,1){5}}
\put(75,60){\vector(-1,1){5}}
\put(70,65){\vector(-1,0){60}}
\put(20,75){\vector(1,0){50}}
\put(70,75){\line(1,1){5}}
\put(75,80){\vector(-1,1){5}}
\put(70,85){\vector(-1,0){40}}
\put(40,0){\vector(0,1){95}}
\put(30,10){\vector(0,1){75}}
\put(20,20){\vector(0,1){55}}
\put(10,30){\vector(0,1){35}}
\put(75,0){\line(0,1){95}}
\put(75,15){\circle*{1}}
\put(75,35){\circle*{1}}
\put(75,60){\circle*{1}}
\put(75,80){\circle*{1}}
\put(34,7){$z$}
\put(32,17){$-z$}
\put(34,27){$z$}
\put(32,37){$-z$}
\put(34,61){$z$}
\put(32,67){$-z$}
\put(34,77){$z$}
\put(32,87){$-z$}
\put(39,96){$z'$}
\put(29,87){$z'$}
\put(19,77){$z'$}
\put(9,67){$z'$}
\put(69,97){${}_i\langle B|$}
\put(70,-3.5){$|B\rangle_i$}
\put(47,0){\vector(0,1){43}}
\put(47,52){\vector(0,1){43}}
\put(46,-2){$z_N$}
\put(46,96){$z'_N$}
\put(46,44){$j_N$}
\put(46,49){$j'_N$}
\put(54,44){$\cdots$}
\put(54,49){$\cdots$}
\put(63,0){\vector(0,1){43}}
\put(63,52){\vector(0,1){43}}
\put(62,-2){$z_1$}
\put(62,96){$z'_1$}
\put(62,44){$j_1$}
\put(62,49){$j'_1$}
\end{picture}
}
\end{picture}

Thanks to (\ref{eq:SWP^}) and (\ref{eq:NWP^}) 
we have 
\begin{equation}
\begin{array}{cl}
&G^{(i)}_N (z, z'| z'_1 , \cdots , z'_N , 
z_N , \cdots , z_1 )^{j'_1 , \cdots , j'_N , j_N , 
\cdots , j_1} \\ 
=&{}_{i}\langle B|A^{(i)}_{SW} (z, z')
\phi_{j'_1} (z'-z'_1 )\cdots \phi_{j'_N} (z'-z'_N )
\phi^{j_N} (z_N -z') \cdots \phi^{j_1} (z_1 -z') 
A^{(i)}_{SW} (z, z')|B\rangle_i . 
\end{array}
\label{df:2N-pt}
\end{equation}
Thus the correlation function $G^{(i)}_N (z, z'| 
z'_1 , \cdots , z'_N , z_N , \cdots , z_1 )^{
j'_1 , \cdots , j'_N , j_N , \cdots , j_1}$ 
normalized by the partition function (\ref{eq:Z}) 
is called the $N$-point local state probability 
of the boundary Belavin model if we set 
$z_l =z'_l =z'=0$, $j_l =j'_l$ ($1\leqq l\leqq N$). 
Owing to the unitarity (\ref{eq:inv-phi}) we have 
\begin{equation}
Z^{(i)}(z_1 , z_2 )=
\sum_{j_1 , \cdots j_N } 
G^{(i)}_N (z, z'| z_1 , \cdots , z_N , 
z_N , \cdots , z_1 )^{j_1 , \cdots , j_N , j_N , 
\cdots , j_1}. 
\end{equation}
Thus we obtain the expression of the $n$-point 
local state probability: 
\begin{equation}
P^{(i)}_N (j_1 , \cdots , j_N )=\frac{
G^{(i)}_N (z, 0| 0, \cdots , 0)^{
j_1 , \cdots , j_N , j_N , \cdots , j_1}}
{\displaystyle\sum_{j_1 , \cdots j_N } 
G^{(i)}_N (z, 0| 0, \cdots , 0)^{
j_1 , \cdots , j_N , j_N , \cdots , j_1}}
\end{equation}

\subsection{Boundary analogue of the quantum 
Knizhnik--Zamolodchikov equation}

Only for $n=2$, the $N$-point function (\ref{df:2N-pt}) 
is reduced 
to the following $2N$-point function of the form 
\begin{equation}
\begin{array}{cl}
&F_{2N}^{(i)}(z, z'| 
y_1 , \cdots , y_N , z_N , \cdots , z_1 )^{
j'_1 , \cdots , j'_N , j_N , \cdots , j_1} \\ 
=& {}_{i}\langle B|A^{(i)}_{SW} (z, z')
\phi^{k_1} (y_1 -z')\cdots 
\phi^{k_N} (y_N -z')
\phi^{j_N} (z_N -z') \cdots \phi^{j_1} (z_1 -z') 
A^{(i)}_{SW} (z, z')|B\rangle_i , 
\end{array}
\label{eq:2N-pt/n=2}
\end{equation}
by putting $y_l =z'_l-w$ and 
$k_l =1-j'_l$ for $1\leqq l \leqq N$ \cite{JKKMW}. 

It is nothing to do with any local state probabilities 
for $n>2$, however, we can consider the correlation 
function of (\ref{eq:2N-pt/n=2})-type: 
\begin{equation}
\begin{array}{rcl}
F^{(i)}_N (z, z'| z_1 , \cdots , z_N )&=&
\displaystyle\sum_{j_1 , \cdots , j_N} v_{j_1} 
\otimes \cdots \otimes v_{j_N} 
F^{(i)}_N (z, z'| z_1 , \cdots , z_N )^{
j_1 , \cdots , j_N} , \\
F^{(i)}_N (z, z'| z_1 , \cdots , z_N )^{
j_1 , \cdots , j_N} &=&
{}_{i}\langle B|A^{(i)}_{SW} (z, z')
\phi^{j_1} (z_1 -z') \cdots \phi^{j_N} (z_N -z') 
A^{(i)}_{SW} (z, z')|B\rangle_i . 
\end{array}
\label{df:N-pt}
\end{equation}
Here we assume that $N\equiv 0$ mod $n$ for simplicity. 

{}From the same discussion as in \cite{Sm1,FR}, we obtain 
\begin{prop} The correlation function (\ref{df:N-pt}) 
satisfies the following relations: 
\begin{eqnarray}
1. \mbox{$R$-matrix symmetry$:$ } 
\hspace{3cm} & & \nonumber \\
P_{j\,j+1}F^{(i)}_N (z, z'| \cdots , z_{j+1} , z_j, \cdots )
&=&R_{j\,j+1}^{V_{z_j }, V_{z_{j+1}}} 
F^{(i)}_N (z, z'| \cdots , z_j , z_{j+1}, \cdots ), 
\label{eq:ax1} \\
2. \mbox{Reflection property $I:$} 
\hspace{2.8cm} & & \nonumber \\
K_N (z_N ) F^{(i)}_N (z, z'| z_1 , \cdots , 
z_{N-1}, z_N ) &=&\nu^{(i)}(z_N ) 
F^{(i)}_N (z, z'| z_1 , \cdots , z_{N-1}, -z_N ), 
\label{eq:ax2-} \\
3. \mbox{Reflection property $I\!I:$} 
\hspace{2.6cm} & & \nonumber \\
\hat{K}_1 (z_1 ) 
F^{(i)}_N (z, z'| z_1 , z_2 , \cdots , z_N ) 
&=&\nu^{(i)}(z_1 ) 
F^{(i)}_N (z, z'| -z_1 -nw, 
z_2 , \cdots , z_N ), 
\label{eq:ax2+}
\end{eqnarray}
where 
$$
\hat{K}(z)v_k =\sum_{j} v_j 
K^{*} (-z-\tfrac{n}{2}w)^k_j . 
$$
\label{prop:ax1,2}
\end{prop}
[Proof] The first equation (\ref{eq:ax1}) follows 
from the commutation relation (\ref{eq:CR-VO}), while 
the second one (\ref{eq:ax2-}) follows from 
(\ref{eq:RE-P}). Finally, from the crossing relation 
(\ref{eq:x-VO}) and (\ref{eq:RE-P_*}) 
$$
\begin{array}{rl}
& \hat{K}_1 (z_1 ) 
F^{(i)}_N (z, z'| z_1 , z_2 , \cdots , z_N ) \\
=&\displaystyle\sum_{j'_1 , j_1 , \cdots , j_N} 
v_{j_1} \otimes \cdots \otimes v_{j_N} 
{}_{i}\langle B|A^{(i+N)}_{SW} (z, z')
\phi^{*}_{j'_1} (z'-z_1 -\tfrac{n}{2}w) 
\cdots 
A^{(i)}_{SW} (z, z')|B\rangle_i 
K^{*}_1 (-z_1 -\tfrac{n}{2}w)^{j'_1}_{j_1} \\
=& \nu^{(i)}(z_1 ) 
\displaystyle\sum_{j_1 , \cdots , j_N} 
v_{j_1} \otimes \cdots \otimes v_{j_N} 
{}_{i}\langle B|A^{(i+N)}_{SW} (z, z')
\phi^{*}_{j_1} (z'+z_1 +\tfrac{n}{2}w) 
\cdots 
A^{(i)}_{SW} (z, z')|B\rangle_i , 
\end{array}
$$
we obtain the last equation (\ref{eq:ax2+}). $\Box$

Owing to the equations 
(\ref{eq:ax1}--\ref{eq:ax2+}) we obtain 
\begin{thm} The correlation function (\ref{df:N-pt}) 
satisfies the following difference equation: 
\begin{equation}
\begin{array}{rcl}
T_j F^{(i)}_N (z, z'| z_1, \cdots , z_N ) 
&=& R_{j j-1}^{V_{z_j -nw},V_{z_{j-1}}}
\cdots R_{j 1}^{V_{z_j -nw},V_{z_{1}}}
\hat{K}_j(-z_j ) \\
&\times&
R_{1 j}^{V_{z_{1},V_{-z_j}}}\cdots
R_{j-1 j}^{V_{z_{j-1},V_{-z_j}}}
R_{j+1 j}^{V_{z_{j+1},V_{-z_j}}} \cdots
R_{N j}^{V_{z_{N},V_{-z_j}}} \\
&\times& K_j (z_j ) 
R_{j N}^{V_{z_j},V_{z_{N}}} \cdots
R_{j j+1}^{V_{z_j},V_{z_{j+1}}}
F^{(i)}_N (z, z'| z_1, \cdots , z_N ), 
\label{eq:N-diff}
\end{array}
\end{equation}
where 
$$
T_j f (z, z'| z_1, \cdots , z_j , 
\cdots , z_N ) =f(z, z'| z_1, \cdots , 
z_j -nw, \cdots , z_N ). 
$$
\end{thm}

Using the crossing symmetries we have another 
expression of the correlation function on the 
dislocated CTM lattice for general $n\geqq 2$: 
\begin{equation}
\begin{array}{cl}
& G^{(i)}_N (z, z'| 
z^*_1 , \cdots , z^*_N , z_N , \cdots , z_1 )^{
j'_1 , \cdots , j'_N , j_N , \cdots , j_1} \\
=&{}_{i}\langle B|A^{(i)}_{SW} (z, z')
\phi^{*j'_1} (z^*_1 -z')\cdots 
\phi^{*j'_N} (z^*_N -z')
\phi^{j_N} (z_N -z') \cdots \phi^{j_1} (z_1 -z') 
A^{(i)}_{SW} (z, z')|B\rangle_i , 
\end{array}
\end{equation}
where $z^*_l =z'_l -\tfrac{n}{2}w$ for $1\leqq l \leqq N$. 
We thus introduce the 
$V^{*\otimes n}\otimes V^{\otimes n}$-valued correlation 
function 
\begin{equation}
\begin{array}{cl}
&G^{(i)}_N (z, z'| z^*_1 , \cdots , z^*_N , 
z_N , \cdots , z_1 ) \\
=&\displaystyle\sum_{j_1 , \cdots , j_N\atop
j'_1 , \cdots j'_n} v^*_{j'_1} 
\otimes \cdots \otimes v^*_{j'_N} \otimes v_{j_N} 
\otimes \cdots \otimes v_{j_1} 
G^{(i)}_N (z, z'| z^*_1 , \cdots , z^*_N , 
z_N , \cdots , z_1 )^{j'_1 , \cdots , j'_N , 
j_N, \cdots , j_1 }. 
\end{array}
\end{equation}

Let us describe the $R$-matrix symmetry corresponding 
to (\ref{eq:ax1}). 
\begin{prop} Let 
\begin{equation}
\begin{array}{cl}
& G^{(\sigma i)}_N (z, z'| x_{\sigma (1)}, 
\cdots , x_{\sigma (2N)}) \\
=& \displaystyle\sum_{j_1 , \cdots , j_N\atop
j'_1 , \cdots j'_n} v^*_{j'_1} 
\otimes \cdots \otimes v^*_{j'_N} \otimes v_{j_N} 
\otimes \cdots \otimes v_{j_1} 
G^{(i)}_N (z, z'| x_{\sigma (1)}, 
\cdots , x_{\sigma (2N)})^{k_{\sigma (1)} , 
\cdots , k_{\sigma (N)}}, \\ 
& G^{(\sigma i)}_N (z, z'| x_{\sigma (1)}, 
\cdots , x_{\sigma (2N)})^{k_{\sigma (1)} , 
\cdots , k_{\sigma (2N)}} \\
=& {}_{i}\langle B|A^{(i)}_{SW} (z, z')
\Phi^{\sigma (1)} \cdots 
\Phi^{\sigma (2N)} 
A^{(i)}_{SW} (z, z')|B\rangle_i . 
\end{array}
\end{equation}
Here $\sigma$ be the permutation of $(1, \cdots , 2N)$, 
and 
$$
x_l =\left\{ \begin{array}{ll} 
z^*_l =z'_l -\tfrac{n}{2}w, & (1\leqq l\leqq N); \\
z_{2N+1-l}, & (N+1 \leqq l\leqq 2N); 
\end{array} \right. 
~~~~
k_l =\left\{ \begin{array}{ll} 
j'_l, & (1\leqq l\leqq N); \\
j_{2N+1-l}, & (N+1 \leqq l\leqq 2N); 
\end{array} \right. 
$$
and 
$$
\Phi^l =\left\{ \begin{array}{ll} 
\phi^{*k_l} (x_l -z'), & (1\leqq l\leqq N); \\
\phi^{k_l}(x_{l}-z'), & (N+1 \leqq l\leqq 2N). 
\end{array} \right. 
$$
Then the following $R$-matrix symmetry holds: 
\begin{equation}
G^{(\sigma_j i)}_N (z, z'|  \cdots , x_{\sigma (j+1)}, 
x_{\sigma (j)}, \cdots ) =
R_{\sigma (j),\sigma (j+1)}^{V^{\sigma (j)},
V^{\sigma (j+1)}}
G^{(\sigma i)}_N 
(z, z'| \cdots , x_{\sigma (j)}, 
x_{\sigma (j+1)}, \cdots ), 
\label{eq:ax1'}
\end{equation}
where 
$$
V^l=\left\{ \begin{array}{ll}
V^*_{x_l} & (1\leqq l \leqq N); \\
V_{x_l} & (N+1\leqq l \leqq 2N); \end{array} 
\right. 
$$
and $\sigma_j$ is the permutation of $(1, \cdots , 2N)$ 
obtained from $\sigma$ by transposing $\sigma (j)$ and 
$\sigma (j+1)$. 
\label{prop:Rsym}
\end{prop}

The reflection properties can be similarly shown 
as before: 
\begin{prop} The following relations holds: 
\begin{eqnarray}
K_{2N} (z_1 ) G^{(\pi\,i)}_N (z, z'| \cdots , z_1 ) 
&=&\nu^{(i)}(z_1 ) 
G^{(\pi\,i)}_N (z, z'| \cdots , -z_1 ), 
\label{eq:axII-} \\
\hat{K}_{2N} (z_1 ) 
G^{(\rho\,i)}_N (z, z'| z_1 , \cdots ) 
&=&\nu^{(i)}(z_1 ) 
T_1 G^{(\rho\,i)}_N (z, z'| -z_1 , \cdots ), 
\label{eq:axII+} \\
\hat{K}^*_1 (z^*_1 ) G^{(\varsigma\,i)}_N 
(z, z'| z^*_1 , \cdots )&=&\nu^{(i)}(-z^*_1 
-\tfrac{n}{2}w)T_1 
G^{(\varsigma\,i)}_N (z, z'| -z^*_1 , \cdots ), 
\label{eq:ax2*-} \\
K^*_{1} (z^*_1 ) G^{(\tau\,i)}_N 
(z, z'|  \cdots , z'_1 )
&=&\nu^{(*i)}(-z^*_1 -\tfrac{n}{2}w)G^{(\tau\,i)}_N 
(z, z'| \cdots , -z^*_1), 
\label{eq:ax2*+}
\end{eqnarray}
Here, 
$$
\hat{K}^* (z) v^*_k =\sum_{j} v^*_j 
K(-z-\tfrac{n}{2}w)^k_j , 
$$
and $\pi$, $\rho$, $\varsigma$, 
$\tau \in \mathfrak{S}_{2N}$ such that 
$$
\pi (2N)=2N, ~~~~ \rho (1)=2N, ~~~~
\varsigma (1)=1, ~~~~ \tau (2N)=1. 
$$
\label{prop:ax1,3}
\end{prop}
[Proof] The relation (\ref{eq:ax1'}) is evident from 
the commutation relations (\ref{eq:CR-VO}). The last 
two (\ref{eq:ax2*-}) and (\ref{eq:ax2*+}) follow from 
(\ref{eq:RE-P*}), (\ref{eq:x-VO}) and (\ref{eq:RE-P_}). 
$\Box$ 

{}From Propositions 
\ref{prop:Rsym} and \ref{prop:ax1,3}, we have 
\begin{thm} Let $V^l_1 =V_{-x_l}$, $V^l_2 =V_{x_l -nw}$. 
Then the following difference equations holds 
\begin{equation}
\begin{array}{rcl}
T_l G_N^{(i)}(z,z'| x_1 , \cdots , x_{2N})
&=&
R_{l\,l-1}^{V^l_2, V^{l-1}}
\cdots R_{l\,1}^{V^l_2, V^1}\hat{K}_l^{*}(-x_l)\\
&\times&
R_{1\,l}^{V^1,V^l_1} \cdots
R_{l-1\,l}^{V^{l-1},V^l_1}
R_{l+1\,l}^{V^{l+1},V^l_1} \cdots
R_{2N\,l}^{V^{2N},V^l_1}
\\
&\times& K_l^{*}(x_l )
R_{l\,2N}^{V^l,V^{2N}}\cdots
R_{l\,l+1}^{V^l,V^{l+1}}
G_N^{(i)}(x_1 , \cdots , x_{2N}),
\end{array}
\label{BQKZ1}
\end{equation}
for $1\leqq l\leqq N$, and 
\begin{equation}
\begin{array}{rcl}
T_l G_N^{(i)}(z,z'| x_1 , \cdots , x_{2N})
&=&
R_{l\,l-1}^{V^l_2, V^{l-1}}
\cdots R_{l\,1}^{V^l_2, V^1}\hat{K}_l(-x_l)\\
&\times&
R_{1\,l}^{V^1,V^l_1} \cdots
R_{l-1\,l}^{V^{l-1},V^l_1}
R_{l+1\,l}^{V^{l+1},V^l_1} \cdots
R_{2N\,l}^{V^{2N},V^l_1}
\\
&\times& K_l (x_l )
R_{l\,2N}^{V^l,V^{2N}}\cdots
R_{l\,l+1}^{V^l,V^{l+1}}
G_N^{(i)}(x_1 , \cdots , x_{2N}),
\end{array}
\label{BQKZ2}
\end{equation}
for $N+1 \leqq l\leqq 2N$. 
\label{thm:LSPn}
\end{thm}

Theorem \ref{thm:LSPn} gives an elliptic 
generalization of the corresponding difference 
equations for the boundary $U_q (\widehat{sl_n})$-symmetric 
model \cite{KQ}. 

\subsection{Boundary spontaneous polarization}

Applying the similar argument as in (\ref{eq:N-diff}) 
to the simplest case $N=1$ we obtain 
the following difference equations: 
\begin{equation}
\begin{array}{rcl}
T_1 G_1^{(i)}(z,z'|z_1^* , z_2 )&=&
\hat{K}^*_1 (-z_1^* ) 
R_{21}^{V_{z_2},V^*_{-z_1^* }} 
K^*_1 (z_1^* ) 
R_{12}^{V^*_{z_1^* },V_{z_2}} 
G_1^{(i)}(z,z'|z_1^* , z_2 ), \\ 
T_2 G_1^{(i)}(z,z'|z_1^* , z_2 )&=&
R_{21}^{V_{z_2 -nw},V^*_{z_1^*}} \hat{K}_2 (-z_2 ) 
R_{12}^{V^*_{z_1^*},V_{-z_2}} K_2 (z_2 ) 
G_1^{(i)}(z,z'|z_1^* , z_2 ), 
\end{array}
\label{eq:d-1pt}
\end{equation}
where $z_1^* =z_1 -\tfrac{n}{2}w$. 
It is difficult to get each element of 
$G_1^{(i)}(z,z'|z_1 , z_2 )$, however, it is 
possible to obtain the expression of the 
following sums: 
\begin{equation}
P^{(i)}_m(z, z'|z_1 , z_2 )=
\displaystyle\sum_{j=0}^{n-1} 
\omega^{mj} G^{(i)}_1 (z, z'|z_1 -\tfrac{n}{2}w, z_2 )^{jj}. 
\label{df:pol0}
\end{equation}
Note that the boundary spontaneous polarization 
as the vacuum expectation value 
of the operator $g$ at boundary is expressed in terms 
of (\ref{df:pol0}) as follows: 
\begin{equation}
\langle g\rangle^{(i)} 
=\left. \dfrac{P^{(i)}_1(z, z'=0|z_1 , z_2 )}{
P^{(i)}_0(z, z'=0|z_1 , z_2 )}\right|_{z_1 =z_2 
=z'}. 
\label{df:pol}
\end{equation}
Now we restrict ourselves to the free boundary condition 
$r\rightarrow 1$ for simplicity. Since 
$\displaystyle\lim_{r\rightarrow 1} {\cal K}(0)
\neq {\cal K}_0$, the initial condition does not 
hold if we take $\overline{K}(z)=
{\cal K}_0 {\cal K}(z)$. Thus we should 
regard the $K$-matrix in this limit as 
$\overline{K}(z)={\cal K}(0) {\cal K}(z)$. Under 
this identification the $K$-matrix behaves as 
$$
K(z)\longrightarrow k(z) I_n , 
$$
where $k(z)$ is a scalar function of $z$. 

Here we cite the following sum formula from 
\cite{SPn}\footnote{
Note that there are typographical errors in the 
formula \cite{SPn}.} 
\begin{equation}
\sum_{j=0}^{n-1} \omega ^{mj} 
\frac{\theta ^{(j)}(z+w)}{\theta ^{(j)}(w)} = 
n 
\frac{h((z-m)/n+w)\prod_{l\neq m} h((-z+l)/n)}
     {h(w)        \prod_{l\neq 0} h(l/n)}, 
\label{eq:th-id}
\end{equation}
Then we see the dual $K$-matrix in the free boundary 
limit $r\rightarrow 1$ behaves as 
$$
K^* (z-\tfrac{n}{2}w) \longrightarrow 
k(-z) f_0 (u^2 q^n )I_n , 
$$
where 
\begin{equation}
\begin{array}{rrl}
f_m (u)&:=&\displaystyle\sum_{j=0}^{n-1} 
\omega ^{mj} R(z)^{j0}_{0j} \\
&=& \dfrac{1}{\bar{\kappa }(u)}
\dfrac{(\omega ^{-m} q^2 u^{-2/n}; t^{2})_{\infty}
      (t^{2}\omega ^{m} q^{-2}u^{2/n}; t^{2})_{\infty}}
     {(\omega ^{m} u^{2/n}; t^{2})_{\infty}
      (t^{2}\omega ^{-m} u^{-2/n}; t^{2})_{\infty}}. 
\end{array} 
\label{df:f_m}
\end{equation}
The difference equations (\ref{eq:d-1pt}) 
are therefore reduced to 
\begin{equation}
\begin{array}{rcl}
T_1 G^{(i)}_1 (z,z'|z^*_1 , z_2 )^{jj}&=&
f_0 (u_1^2 q^n )
\displaystyle\sum_{k,l} R_{12}(-z_1 -z_2 )^{kj}_{jk}
R_{21}(z_2 -z_1 )_{kl}^{lk} 
G^{(i)}_1 (z,z'|z_1^* , z_2 )^{ll}, \\
T_2 G^{(i)}_1 (z,z'|z_1^* , z_2 )^{jj}&=&
f_0 (u_2^2 q^n )
\displaystyle\sum_{k,l} R_{12} (z_1 -z_2 )^{kj}_{jk}
R_{21}(-z_1 -z_2 )_{kl}^{lk} 
G^{(i)}_1 (z,z'|z_1^* , z_2 )^{ll}, 
\end{array}
\label{eq:1p-red}
\end{equation}
where $z_1^* =z_1 -\tfrac{n}{2}w$, and 
we use (\ref{eq:cross}) and (\ref{eq:K-inv}). 
Substituting (\ref{eq:1p-red}) into (\ref{df:pol0}) we obtain 
\begin{equation}
P^{(i)}_m (z,z'|z_1 , z_2 )=C^{(i)}_m A(u_1 )A(u_2 )
B_m (u_+ )B_{-m}(u_- ). 
\label{eq:sol-P}
\end{equation}
Here $C^{(i)}_m$ is a constant, 
and $A(u)$ and $B_m (u)$ are solutions to the 
following difference equations:  
\begin{equation}
\dfrac{A(uq^{n})}{A(u)}=f_0 (u^2 q^n ), ~~~~
\dfrac{B_m (uq^{-n})}{B_m (u)}=f_m (u). 
\end{equation}
By solving these difference equations we obtain 
\begin{equation}
A(u)=\psi (u^2 )
\dfrac{(q^2 u^{4/n}; t^2 , q^{4})_{\infty}
      (q^{4}u^{-4/n}; t^2 , q^{4})_{\infty}}
      {(t^2 u^{4/n}; t^{2}, q^4 )_{\infty}
      (t^{2}q^2 u^{-4/n}; t^{2}, q^4 )_{\infty}}, 
\end{equation}
where 
$$
\psi (u):=g_0 (uq^{-n/2})g(u^{-1}q^{n/2}), ~~~~
g_0 (u):=\frac{(q^{2+3n} u^{-2}; t^2 , q^{2n}, q^{4n})_{\infty} 
            (t^2 q^{-2+3n} u^{-2}; t^2 , q^{2n}, q^{4n})_{\infty}} 
           {(q^{3n} u^{2}; t^2 , q^{2n}, q^{4n})_{\infty} 
            (t^2 q^{3n} u^{2}; t^2 , q^{2n}, q^{4n})_{\infty}}; 
$$
and 
\begin{equation}
B_m (u)=\varphi (u)
\dfrac{(t^2 \omega^{m} u^{2/n}; t^{2})_{\infty}
      (t^{2}\omega^{-m} u^{-2/n}; t^{2})_{\infty}}
     {(q^2 \omega^{m} u^{2/n}; q^{2})_{\infty}
      (q^{2}\omega^{-m}u^{-2/n}; q^{2})_{\infty}}, 
\label{eq:sol-B}
\end{equation}
where 
$$
\varphi (u):=g(uq^{n/2})g(u^{-1}q^{n/2}), ~~~~
g(u):=\frac{(q^{3n} u^{-2}; t^2 , q^{2n}, q^{2n})_{\infty} 
            (t^2 q^{3n} u^{-2}; t^2 , q^{2n}, q^{2n})_{\infty}} 
           {(q^{2+n} u^{2}; t^2 , q^{2n}, q^{2n})_{\infty} 
            (t^2 q^{-2+n} u^{2}; t^2 , q^{2n}, q^{2n})_{\infty}}. 
$$
Note that $B_m (u)$ is essentially the same as 
$G^{(m)}(u)$ in \cite{SPn}, which corresponds to 
the quantity (\ref{df:pol0}) in the bulk theory. 

{}From (\ref{eq:sol-P}) we have 
\begin{eqnarray}
\dfrac{P^{(i)}_1 (z, z'=0|z_1 , z_2 )}{
P^{(i)}_0(z, z'=0|z_1 , z_2 )}
&=&\dfrac{C^{(i)}_1}{C^{(i)}_0} 
\dfrac{B_1 (u_+ ) B_{-1}(u_- )}
     {B_0 (u_+ )B_0 (u_- )}. 
\label{eq:Eu} 
\end{eqnarray}
Taking the low temperature limit $t,q\rightarrow 0$, 
we find that the ratio $C^{(i)}/C^{(i)}_0$ should 
be equal to $\omega^i$. We therefore obtain the 
boundary spontaneous polarization from (\ref{eq:Eu}) 
and (\ref{eq:sol-B}) by putting $u_+=u_- =1$ 
\begin{equation}
\langle g \rangle ^{(i)}=\omega^i 
\displaystyle\frac{(q^2 ; q^{2})_{\infty}^{4}} 
                  {(t^2 ; t^{2})_{\infty}^{4}} 
\displaystyle\frac{(t^2 \omega ; t^{2})_{\infty}^2
                   (t^{2}\omega ^{-1} ; t^{4})_{\infty}^2}
                  {(q^2 \omega ; q^{2})_{\infty}^2
                   (q^{2}\omega ^{-1} ; q^{4})_{\infty}^2}.
\end{equation}
When $n=2$ this expression coincides with the 
previous result obtained in \cite{JKKMW}. We also 
emphasize that the boundary spontaneous polarization 
for the boundary Belavin model is exactly the square 
of that for the bulk Belavin model obtained in 
\cite{SPn}, up to a phase factor. 

\section{Summary and discussion}

In this paper we have obtained two non-diagonal solutions 
of the reflection equation associated with Belavin's 
$\bz_n$-symmetric elliptic model. Unfortunately, our 
elliptic $K$-matrix is not connected with the diagonal 
boundary Boltzmann weights for the $A^{(1)}_{n-1}$-face 
model \cite{BFKZ} but the non-diagonal ones. 
It is thus an open problem to obtain the 
$K$-matrix corresponding to the boundary Boltzmann 
weights given in \cite{BFKZ}. 

On the basis of the boundary CTM bootstrap 
we have derived a set of difference 
equations for correlation functions of the boundary 
Belavin model. By solving the simplest difference 
equations, we have obtained 
the boundary spontaneous polarization of the boundary 
Belavin model. Our result is consistent with the one 
given in \cite{JKKMW} when $n=2$. The boundary 
spontaneous polarization is equal to the square of 
the bulk spontaneous polarization \cite{SPn} up to 
a phase factor. The same phenomena were observed in 
\cite{JKKKM,JKKMW}. 

In this paper we have shown that correlation functions 
of the boundary model satisfy the $R$-matrix symmetry and 
the reflection properties, which are the boundary analogue 
of Smirnov's first two axioms \cite{S}. It may be 
interesting to construct integral formulae for correlation 
functions such that the integrand possesses the determinant 
structure as in Smirnov's integral \cite{S}. 

In \cite{H} integral formulae for correlation functions 
of the boundary $XYZ$ model by using bosonization of 
vertex operators \cite{LP}. In order to obtain the 
higher $n$ generalization of \cite{H}, the construction 
of free field realization of the boundary 
Belavin model is required. It is a very hard but important 
work. 

\section*{Acknowledgement}

The author would like to thank A. Kuniba and 
A. Nakayashiki for discussion.

\end{document}